\documentclass[aps,prl,twocolumn,superscriptaddress,preprintnumbers,%
               showpacs,nofootinbib]{revtex4}
\newcommand{\PRE}[1]{}       

\usepackage{bm}
\usepackage{epsfig}

\newcommand{\postscript}[2]{\setlength{\epsfxsize}{#2\hsize}
   \centerline{\epsfbox{#1}}}

\def\eslt{\not\!\!{E_T}}
\def\to{\rightarrow}

\def\bi{\begin{itemize}}
\def\ei{\end{itemize}}
\def\te{\tilde e}

\def\ta{\tilde a}

\def\tb{\tilde b}

\def\tst{\tilde t}

\def\tg{\tilde g}

\def\tell{\tilde\ell}
\def\tq{\tilde q}
\def\tw{\widetilde W}
\def\tz{\widetilde Z}
\def\alt{\stackrel{<}{\sim}}
\def\agt{\stackrel{>}{\sim}}
\def\be{\begin{equation}}  
\def\ee{\end{equation}}  
\def\bea{\begin{eqnarray}}  
\def\eea{\end{eqnarray}}  
\newcommand\plb[3]{{Phys.\ Lett.\ }{\bf B #1} (#2) #3}
\newcommand\jhep[3]{{J. High Energy Phys.\ }{\bf #1} (#2) #3}

\newcommand\npb[3]{{Nucl.\ Phys.\ }{\bf B #1} (#2) #3}

\begin{document}

\preprint{OU-HEP-150214}

\title{
\PRE{\vspace*{1.5in}}
Supergravity gauge theories strike back:\\
There is no crisis for SUSY but a new collider may be required for discovery
\PRE{\vspace*{0.3in}}
}

\author{Howard Baer}
\affiliation{Dept. of Physics and Astronomy,
University of Oklahoma, Norman, OK, 73019, USA
\PRE{\vspace*{.1in}}
}
\author{Vernon Barger}
\affiliation{Dept. of Physics,
University of Wisconsin, Madison, WI 53706, USA
\PRE{\vspace*{.1in}}
}
\author{Mike Savoy}
\affiliation{Dept. of Physics and Astronomy,
University of Oklahoma, Norman, OK, 73019, USA
\PRE{\vspace*{.1in}}
}


\begin{abstract}
\PRE{\vspace*{.1in}} 
More than 30 years ago, Arnowitt-Chamseddine-Nath (ACN) and others
established the compelling framework of supergravity gauge theories (SUGRA) 
as a picture for the next step in beyond the Standard Model physics.
We review the current SUGRA scenario in light of recent data from
LHC8 collider searches and the Higgs boson discovery.
While many SUSY and non-SUSY scenarios are highly disfavored or even excluded
by LHC, the essential SUGRA scenario remains intact and as compelling as
ever. For naturalness, some non-universality between matter and Higgs sector
soft terms is required along with substantial trilinear soft terms.
SUSY models with radiatively-driven naturalness (RNS) are found with 
high scale fine-tuning at a modest $\sim 10\%$. 
In this case, natural SUSY might be discovered at LHC13 but could also 
easily elude sparticle search endeavors. 
A linear $e^+e^-$ collider with $\sqrt{s}>2m(higgsino)$
is needed to provide the definitive search for the required light higgsino
states which are the hallmark of natural SUSY.
In the most conservative scenario, we advocate inclusion of a Peccei-Quinn
sector so that dark matter is composed of a WIMP/axion admixture
{\it i.e.} two dark matter particles.

\end{abstract}

\pacs{12.60.-i, 95.35.+d, 14.80.Ly, 11.30.Pb}

\maketitle

\section{Introduction}

The recent amazing discovery of a Higgs scalar with mass
$m_h\simeq 125$ GeV by the Atlas\cite{atlas_h} and CMS\cite{cms_h} collaborations 
at LHC seemingly completes the Standard Model (SM), 
and yet brings with it a puzzle.
It was emphasized as early as 1978 by Wilson/Susskind\cite{techni1} 
that fundamental scalar particles are unnatural in quantum field theory. 
In the case of the SM Higgs boson with a doublet of Higgs scalars $\phi$ 
and Higgs potential given by
\be 
V=-\mu^2\phi^\dagger\phi +\lambda (\phi^\dagger\phi )^2 ,
\ee
one expects a physical Higgs boson mass value
\be
m_h^2\simeq 2\mu^2 +\delta m_h^2
\label{eq:mhSM}
\ee 
where the leading radiative correction is given by
\be
\delta m_h^2\simeq \frac{3}{4\pi^2}\left(-\lambda_t^2+\frac{g^2}{4}+
\frac{g^2}{8\cos^2\theta_W} +\lambda \right)\Lambda^2 .
\ee
In the above expression, $\lambda_t$ is the top quark Yukawa coupling, 
$g$ is the $SU(2)$ gauge coupling and $\lambda$ is the Higgs field 
quartic coupling. The quantity $\Lambda$ is the UV energy cutoff to 
otherwise divergent loop integrals. Taking $\Lambda$ as high as
the reduced Planck mass $M_P\simeq 2.4\times 10^{18}$ GeV would require
a tuning of $\mu^2$ to $30$ decimal places to maintain the
measured value of $m_h^2$. Alternatively, naturalness--
requiring that no parameter needs to be adjusted to unreasonable accuracy
(as articulated by Dimopoulos and Susskind\cite{etc2})--
required that loop integrals be truncated at $\Lambda\sim 1$ TeV: 
{\it i.e.} one expects the SM to occur as an effective field theory valid 
below $\sim 1$ TeV, and that at higher energies new degrees 
of freedom will be required. While the technicolor route{\cite{etc1,etc2} 
banished all fundamental scalars from the theory, 
an attractive alternative which
naturally admitted fundamental scalars-- supersymmetry-- was already emerging.

In supersymmetry, the fundamental bose-fermi spacetime symmetry
guaranteed cancellation of all quadratic scalar mass divergences
so that scalar fields could co-exist with their well-behaved
fermion and gauge-boson brethren.
Early models based on global SUSY could be seen to lead to phenomenological 
inconsistencies: some superpartners would have to exist with masses below
their SM partners: such a situation-- {\it e.g.} the presence of scalar 
electrons with mass less than an electron-- 
would not have eluded experimental detection.
The simultaneous development of models based on gauged, or local SUSY, 
provided a path forward which was consistent with phenomenological requirements.
Local SUSY models-- where the spinorial SUSY transformation parameter $\alpha$
in $e^{-i\bar{\alpha}Q}$ depended explicitly in spacetime $\alpha(x)$--
required the introduction of a gravitino-graviton supermultiplet, and hence
were called supergravity theories, or SUGRA for short\cite{an,superg}. 
The SUGRA sum rules for
sparticle masses were modified so that all the unseen superpartner masses 
could be lifted up to the fundamental scale set by the gravitino mass $m_{3/2}$.
Since these theories necessarily included gravity, they also necessarily 
contained non-renormalizable terms multiplied by powers of $1/M_P$.
The modern viewpoint is then that SUGRA theories might be the low energy 
effective theory obtained from some more fundamental ultra-violet
complete theory such as superstrings.

\subsection{SUGRA gauge theories}

The starting point for construction of realistic supersymmetric models
was the development of the Lagrangian for $N=1$ locally supersymmetric gauge 
theories. The final result, obtained by Cremmer {\it et al.} in 
1982\cite{Cremmer:1982en}
is now textbook material\cite{wessbagger,wss}. The locally supersymmetric
Lagrangian for SUSY gauge theories-- after elimination of all auxiliary fields 
and in four-component notation with a $+,-,-,-$ 
metric-- is written down over several pages in \cite{wss}.

To construct SUGRA gauge theories\cite{sugra}\footnote{For an historical review, see {\it e.g.} Ref.~\cite{acnreview}.}, 
a multi-step procedure can be followed:
\begin{enumerate}
\item stipulate the desired gauge symmetry for the theory,
\item stipulate the super-field content consisting of chiral scalar
superfields (containing spin-$1/2$ matter and spin-0 superpartners),
the appropriate gauge superfields in accord with the gauge symmetry
from step 1 (these contain massless gauge bosons and spin-$1/2$ gauginos) and
the graviton-gravitino supermultiplet,
\item the remaining model freedom comes from stipulating the form of 
the holomorphic gauge kinetic function $f_{AB}(\phi_i )$ 
and superpotential $W(\phi_i )$ and the real Kahler potential
$K(\phi^\dagger ,\phi )$. In SUGRA theories, the K\"ahler potential
and the superpotential necessarily are combined into the K\"ahler
function $G=K/M_P^2+\log|W/M_P^3|^2$.
\end{enumerate}

In complete analogy to the Higgs mechanism in local gauge theories, 
SUGRA theories allow for a superHiggs mechanism.
In the superHiggs mechanism, if one arranges for a breakdown in local
SUSY, then instead of generating a physical goldstino field, 
the spin-$1/2$ goldstino is eaten by the spin-$3/2$ gravitino so that 
the gravitino gains a mass $m_{3/2}$.
To accomodate a breakdown in SUGRA, it is necessary to introduce a
``hidden sector'' of fields $h_m$. The hidden sector serves as an arena
for SUSY breaking. 

Many early models invoked a very simple SUSY breaking sector. The
fields were divided between an observable sector $C_i$ and 
a hidden sector $h_m$ with a separable superpotential
$W=W_{obs}(C_i)+W_{hidden}(h_m)$ and a flat K\"ahler
metric: $K=C_i^\dagger C_i+h_m^\dagger h_m$.
A single hidden sector field $h$ might obey the Polonyi superpotential:
\be
W_{Polonyi}=m_{hidden}^2(h+\beta) .
\ee
The $F$-type SUSY breaking condition 
$\frac{\partial{W}}{\partial h}+\frac{h^*W}{M_P^2}\ne 0$ is satisfied 
for $\beta^2<4M_P^2$. 
While the $F$-term of $h$ gains a VEV 
$\langle F_h\rangle\sim m_{hidden}^2$, 
the scalar component gains a VEV $\langle h\rangle\sim M_P$. 
The gravitino becomes massive
\be
\frac{i}{2}e^{G/2}M_P \bar{\psi}_\mu \sigma^{\mu\nu}\psi_\nu\to 
\frac{i}{2}e^{G_0/2}M_P\bar{\psi}_\mu\sigma^{\mu\nu}\psi_\nu
\ee
where $G_0$ is the VEV of $G$. The gravitino mass is given by
\be
m_{3/2}=e^{G_0/2}M_P\sim m_{hidden}^2/M_P .
\ee 
A TeV value of $m_{3/2}$ is achieved for a hidden sector mass scale 
$m_{hidden}\sim 10^{11}$ GeV.

Once the gravitino gains mass, then an amazing simplification occurs.
By replacing the hidden sector fields by their VEVs and taking the 
flat space limit
$M_P\to\infty$ while keeping $m_{3/2}$ fixed, one arrives at the 
Lagrangian of global SUSY for the visible sector fields 
augmented by {\it soft SUSY breaking terms}
consisting of gaugino masses $M$ (assuming a non-trivial form for 
the gauge kinetic function), scalar squared masses $m_{\phi_i}^2$, 
trilinear $a$ and bilinear $b$ soft terms\cite{sugra}.
The soft terms all turn out to be multiples of the gravitino mass 
$m_{3/2}$. For the case of the Polonyi model, then one expects
\bea
m_{\phi_i}^2& =&m_{3/2}^2,\nonumber \\
A&=& (3-\sqrt{3})m_{3/2},\nonumber \\
B&=& A-m_{3/2}=(2-\sqrt{3})m_{3/2}\ {\rm while}\nonumber \\
M&\sim & m_{3/2}
\label{eq:polonyi}
\eea
until one specifies additionally the gauge kinetic function.
The universality of scalar masses and trilinears is welcome in that
it allows for the super-GIM mechanism to suppress flavor violating
processes while the reality of soft terms suppresses unwanted CP violation. 

While the Polonyi model soft term values are intriguing, ordinarily one does
not take such a toy model seriously as being indicative of the hidden sector.
More general expressions for the soft terms for a general hidden sector,
including a non-flat K\"ahler metric, have been calculated in 
Ref's \cite{Soni:1983rm,kl,Brignole:1993dj}. The result is that: under a well-specified 
hidden sector, the soft SUSY breaking terms still arise as multiples 
of $m_{3/2}$ although universality is not assured so that, in general, one expects
both flavor and CP-violating processes to occur. Experimental limits
on such processes provide constraints to SUGRA model building efforts.

In general, there may occur a multitude of hidden sector fields along
with additional hidden sector gauge symmetries. In $4-D$ string theory, 
an automatic hidden sector can arise in the form of the 
dilaton field $S$ and the moduli fields $T_m$ that parametrize the 
size and shape of the compactification of the extra dimensions.
In addition, if there are additional hidden sector gauge groups--
as would arise in $E_8\times E_8^\prime$ heterotic string theory--
and if the additional gauge forces become strong at an intermediate scale
$\Lambda\sim 10^{13}$ GeV, then hidden sector gauginos may condense\cite{gaugino} 
resulting in a breakdown of SUSY with $m_{3/2}\sim \Lambda^3/M_P^2$.

In spite of the daunting plethora of hidden sector possibilities, 
it is still possible to make progress in matching theory to experiment
by appealing to {\it effective field theories}. In spite of our lack of 
knowledge of hidden sector dynamics, we may {\it parametrize our ignorance}
by largely eschewing the hidden sector altogether and replacing it 
by {\it an adjustable set of soft SUSY breaking parameters}.
As we scan over various soft parameter values, then we are
effectively accounting for a wide variety of hidden sector possibilities.
Under this plan, it is possible to make additional assumptions as to how the 
various soft terms are related to one another. For instance, one might
assume universality to suppress FCNC and CP violating processes, 
or one might assume various GUT relations or relations amongst 
soft terms arising from different string theory possibilities.

\subsection{Connection to weak scale supersymmetry}

It is usually assumed that the induced soft SUSY breaking terms arise at 
or around the reduced Planck scale $M_P$. Their values at lower energy 
scales are obtained by solving their renormalization group equations 
(RGEs)\cite{RGEs,rge1992}. 
Inspired by 1. the fact that gauge couplings unify at a
scale $m_{GUT}\simeq 2\times 10^{16}$ GeV, and 2. that the most parsimonius
effective theory below the GUT scale is the 
minimal supersymmetric standard model (MSSM), 
the soft terms are usually imposed at $m_{GUT}$ where it may be 
understood that some above-the-GUT-scale
running may have already occured, perhaps in the context of some actual
GUT construct\cite{Dimopoulos:1981zb,polonsky,bdqt}.

Under the assumption that the gaugino masses unify at $m_{GUT}$
(as they ought to if some simple GUT holds above $m_{GUT}$ or if 
the gauge kinetic function has a universal dependence on hidden sector fields)
then we expect
\bea
m(bino)&\equiv &M_1 \sim 0.44 m_{1/2},\\
m(wino)&\equiv &M_2\sim 0.81 m_{1/2},\\
m(gluino)&\equiv &M_3\sim 2.6 m_{1/2}, 
\eea
where $m_{1/2}$ is the unified gaugino mass at $Q=m_{GUT}$.
The electroweak gauginos mix with the higgsinos to yield
two physical charginos $\tw_{1,2}^\pm$ and four neutralinos $\tz_{1,2,3,4}$
ordered according to ascending mass.
Also, the weak scale values of the squark and slepton masses
are given by
\bea
m_{\tq}^2 &\simeq & m_0^2 + (5-6)m_{1/2}^2\;,  \\
m_{\te_L}^2 &\simeq & m_0^2 + 0.5 m_{1/2}^2\;, \\
m_{\te_R}^2 &\simeq & m_0^2 + 0.15m_{1/2}^2\; ,
\eea
where $m_0$ is the unified scalar mass at $Q=m_{GUT}$.
For more precise values, including mixing effects and 
radiative corrections\cite{pbmz}, one may consult one of several
computer codes available for SUSY mass spectra\cite{isajet,Allanach:2003jw}.

A potentially tragic feature of this construct is that the
soft terms which enter the scalar (Higgs) potential, 
$m_{H_u}^2$ and $m_{H_d}^2$, are $\sim m_{3/2}^2$ and so 
manifestly positive. But phenomenology dictates that the scalar
potential should develop a non-zero minimum so that 
electroweak symmetry is properly broken: $SU(2)_L\times U(1)_Y\to U(1)_{EM}$.
It was conjectured already in 1982 that, if the top quark mass was 
large enough, then radiative effects could drive exactly the right soft
term $m_{H_u}^2$ to negative values so that EW symmetry is properly
broken. This radiative electroweak symmetry breaking (REWSB) could occur
if the top quark mass lay in the 100-200 GeV range\cite{rewsb}. 
While such a heavy top quark seemed crazy at the time, the ultimate discovery of the top
quark with mass $m_t=173.2\pm 0.9$ GeV has vindicated this approach.

A final oddity in the SUGRA gauge theory construct is the allowance of a 
mass term in the superpotential: $W_{MSSM}\ni \mu H_u H_d$. 
Since this term is supersymmetric and not SUSY breaking, one would expect
it to occur with a value $\mu\sim M_P$. However, for an appropriate 
breakdown of electroweak symmetry and to naturally develop a weak
scale VEV, then $\mu$ is required to be $\sim M_Z$.

There are several approaches to this so-called SUSY $\mu$ problem.
All require as a first step the imposition of some symmetry to forbid 
the appearance of $\mu$ in the first place. For instance, if the Higgs
multiplets carry Peccei-Quinn charges, then $\mu$ is forbidden
under the same PQ symmetry which is also needed to solve the strong CP problem.
Next, one introduces extra fields to couple to the Higgs multiplets.
Invoking hidden sector field(s) which couple to $H_uH_d$ in the K\"ahler
potential via non-renormalizable operators
\be
K\ni \lambda h^\dagger H_u H_d/M_P
\ee
(where the $F$-term of $h$ develops a VEV $\langle F_h\rangle\sim m_{hidden}^2$) 
leads to a $\mu$ term
\be
\mu\sim \lambda m_{hidden}^2/M_P
\ee 
which is of order $m_Z$ for $m_{hidden}\sim 10^{11}$ GeV. 
This is the Giudice-Masiero mechanism\cite{GM}. 
Alternatively, coupling the Higgs fields to a visible sector
singlet $W_{NMSSM} \ni \lambda S H_uH_d$, where $\phi_S$ develops a weak scale
VEV, then leads to the Next-to-Minimal Supersymmetric Standard Model or NMSSM\cite{nmssm}.
A third possibility-- Kim-Nilles\cite{KN}, which includes the PQ strong $CP$ solution in an intimate way-- 
is to couple the Higgs fields to a PQ superfield $S$ so that 
$W_{DFSZ}\ni\lambda S^2H_uH_d/M_P$. 
This is the supersymmetrized version of the DFSZ axion model and leads to a $\mu$ term,
$\mu\sim \lambda f_a^2/M_P$, which gives $\mu\sim M_Z$ for
an axion decay constant $f_a\sim m_{hidden}\sim 10^{10}$ GeV.

\section{Status of Sugra gauge theories}

There are three indirect experimental success stories for supersymmetric models.
These are indirect in that they do not involve direct confirmation of weak scale SUSY by 
detection of supersymmetric matter (which would be the most important way to confirm SUSY), 
but instead they each involve virtual contributions
of supersymmetric matter to experimental observables. Had any of these three measurements 
turned out quite differently, then supersymmetric models would have been placed in a difficult-- 
perhaps untenable-- position.

\subsection{Experimental successes}

\subsubsection{Gauge coupling unification}

The measurements of the three SM gauge couplings to high precision over the years-- 
especially from measurements at LEP2, Tevatron and LHC-- 
have provided perhaps the most impressive experimental support for SUSY.
From the measured values of the Fermi constant $G_F$, the $Z$-boson mass $m_Z$, the electromagnetic coupling
$\alpha_{EM}$ and the top quark mass $m_t$, the $U(1)_Y$ and $SU(2)_L$ gauge couplings $g_1$ and 
$g_2$ can be computed at scale $Q=M_Z$ in the $\overline{DR}$ regularization scheme. 
A variety of measurements also constrain the value of $\alpha_s\equiv g_3^2/{4\pi}$ at 
$Q=M_Z$. These serve as weak scale inputs to test whether the gauge couplings actually do unify
as expected in a GUT theory or not. 
For gauge coupling RGEs in the MSSM, the couplings do indeed unify to a precision of about 
a few percent; in contrast, for the SM or MSSM augmented by extra non-GUT matter, then the unification
fails utterly.

\subsubsection{Top quark mass and electroweak symmetry breaking}

As mentioned previously, the Higgs potential with soft scalar masses $m_\phi^2\sim m_{3/2}^2$ 
does not admit the non-zero Higgs VEV which is needed for electroweak symmetry breaking.
However, the expected value $m_{H_u,d}^2\sim m_{3/2}^2$ is imposed at some high scale
such as $Q=M_P$ or $m_{GUT}$ and is modified by radiative corrections. An appropriate
EWSB is obtained if $m_{H_u}^2$ runs to negative values at the weak scale. 
The relevant RGE is given by
\be
\frac{dm_{H_u}^2}{dt}=\frac{2}{16\pi^2}\left(-\frac{3}{5}g_1^2M_1^2-3g_2^2M_2^2+\frac{3}{10}g_1^2S
+3f_t^2X_t\right)
\label{eq:dmHu}
\ee
with $X_t=m_{Q_3}^2+m_{U_3}^2+m_{H_u}^2+A_t^2$ and where $S=0$ for models with universal scalars. 
While the gauge terms in Eq. \ref{eq:dmHu} push $m_{H_u}^2$ to larger values as $t=\ln (Q^2)$ runs from
$m_{GUT}$ to $m_{weak}$, the term involving the top quark Yukawa coupling $f_t$ pushes
$m_{H_u}^2$ towards negative values. The top-Yukawa term typically wins out for
the top quark mass $m_t\sim 100-200$ GeV. Had the value of $m_t$ been found to be below
$\sim 100$ GeV, then EWSB would be hard pressed in SUSY and other exotica would have been required.
The situation is shown in Fig. \ref{fig:mtmh} where we show the shaded regions of the $m_t$ vs. $m_0$
plane where EWSB successfully occurs. For this case, we choose a mSUGRA/CMSSM model
benchmark with $m_{1/2}=700$ GeV, $A_0=-1.6m_0$ and $\tan\beta =10$. We also show contours of
light Higgs mass $m_h$. In this case, a Higgs mass $m_h\sim 125$ GeV is achieved for $m_t\sim 175$ GeV.
\begin{figure}[tbp]
\postscript{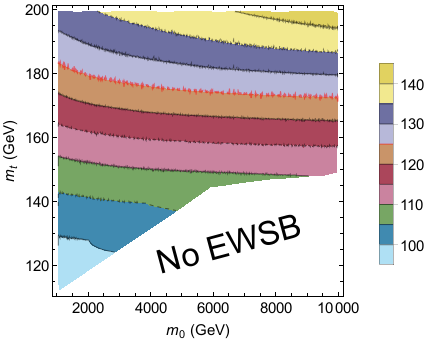}{0.95}
\caption{
Values of Higgs mass $m_h$ which are generated for various values of
$m_t$ and $m_0$ in a SUSY model with $m_{1/2}=700$ GeV and
$A_0=-1.6 m_0$ and $\tan\beta =10$.
\label{fig:mtmh}}
\end{figure}

\subsubsection{The mass of the Higgs boson}

In the Standard Model, the mass of the Higgs boson is given by Eq. \ref{eq:mhSM}. 
Prior to discovery, its
mass could plausibly lie anywhere from the lower limit established by LEP2 searchs-- 
$m_h>114.1$ GeV-- up to $\sqrt{8\pi\sqrt{2} /3 G_F}\sim 800$ GeV as 
required by unitarity\cite{Lee:1977eg}.
This mass range is exhibited in Fig. \ref{fig:mh} as the blue band.
\begin{figure}[tbp]
\postscript{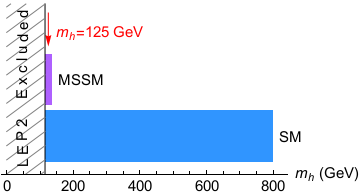}{0.95}
\caption{
Range of Higgs mass $m_h$ predicted in the Standard Model compared to range
of Higgs mass predicted by the MSSM.
We also show the measured value of the Higgs mass by the arrow.
The left-most region had been excluded by LEP2 searches prior to 
the LHC8 run.
\label{fig:mh}}
\end{figure}

In contrast, in the MSSM the Higgs mass is calculated at the 1-loop
level as\cite{mhiggs}
\be
m_h^2\simeq M_Z^2\cos^2 2\beta +\frac{3g^2}{8\pi^2}\frac{m_t^4}{m_W^2}\left[
\ln\frac{m_{\tst}^2}{m_t^2}+\frac{X_t^2}{m_{\tst}^2}\left(1-\frac{X_t^2}{12m_{\tst}^2}\right)\right]
\label{eq:mhiggs}
\ee
where $X_t=A_t-\mu\cot\beta$ and $m_{\tst}^2\simeq m_{Q_3}m_{U_3}$ is an effective squared stop mass.  
For a given $m_{\tst}^2$, this expression is maximal for large mixing in the
top-squark sector with $X_t^{max}=\sqrt{6}m_{\tst}$ (see Fig. \ref{fig:azar}). 
\begin{figure}[tbp]
\postscript{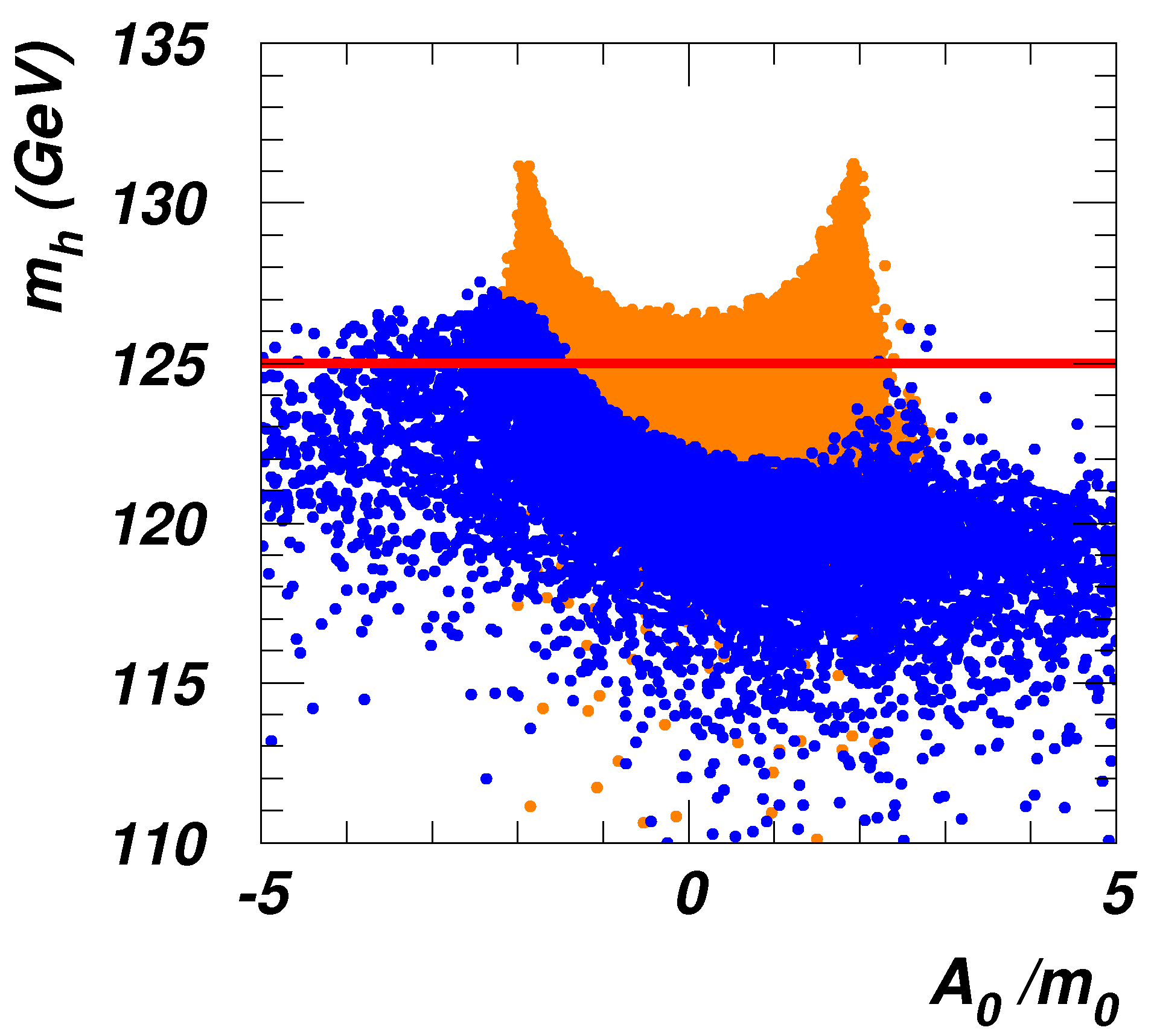}{0.95}
\caption{
Predicted mass of the light Higgs scalar $m_h$ from a scan over
NUHM2 SUSY model parameter space with $m_0$ ranging up to 5 TeV
(blue points) and up to 20 TeV (orange points)
taken from Ref. \cite{h125}.
\label{fig:azar}}
\end{figure}

For top-squark masses not much beyond the 
TeV scale, the upper limit on the SM-like SUSY Higgs boson is $m_h\alt 135$ GeV.
This range is shown as the purple band. 
The combined Atlas/CMS measured value of the newly discovered 
Higgs boson is given by
\be
m_h= 125.15\pm 0.24\ {\rm GeV}
\ee
and is indicated by the red arrow in Fig. \ref{fig:mh}.

\subsection{Collider searches for supersymmetric matter}

While SUSY models enjoy compatibility with the measured values of the gauge couplings,  
the top quark mass and Higgs boson mass, the main goal is to discover supersymmetry via 
the direct detection of supersymmetric matter at colliding beam experiments. 
The CERN LEP2 $e^+e^-$ collider searched for SUSY in various guises without success. 
The most important bound to emerge from LEP2 was that chargino masses $m_{\tw_1}>103.5$ GeV
in a relatively model-independent way as long as the mass gap $m_{\tw_1}-m_{\tz_1}$ is
greater than just several GeV.

At the CERN LHC, a variety of searches for SUSY particle production have taken place
at $pp$ collisions at $\sqrt{s}=8$ TeV. 
For sparticle masses in the TeV regime, the most lucrative production channel-- owing to large
cross sections followed by expected large energy release in cascade decays\cite{cascade}-- 
is gluino and squark pair production: $pp\to\tg\tg,\ \tq\tq$ and $\tg\tq$. 
From these processes, a variety of multi-jet plus multi-lepton plus $\eslt$ events are
expected\cite{bcpt} provided the sparticle masses are light enough that production cross sections
are sufficiently large. So far, no compelling signal has been seen above expected background 
levels\cite{atlas_susy,cms_susy}. The resulting excluded regions of SUSY parameter space are shown in Fig. \ref{fig:atlas}
in the context of the mSUGRA model with $\tan\beta =30$ and $A_0=-2m_0$ (values which
ensure a Higgs mass $m_h\sim 125$ GeV throughout much of the parameter space shown).
The left side of the plot shows the region where squark and gluino masses are comparable
$m_{\tq}\sim m_{\tg}$ so that $\tg\tg$, $\tq\tq$ and $\tg\tq$ can all occur at comparable rates.
The right side of the plot shows the region where $m_{\tq}\gg m_{\tg}$ so that only 
$\tg\tg$ production is relevant. From the plot, we can read off the approximate bounds:
\bea
m_{\tg}&\agt &1300\ {\rm GeV}\ \ \ \ (m_{\tg}\ll m_{\tq})\ {\rm and}\\
m_{\tg}&\agt &1800\ {\rm GeV}\ \ \ \ (m_{\tg}\sim m_{\tq}) .
\eea
\begin{figure}[tbp]
\postscript{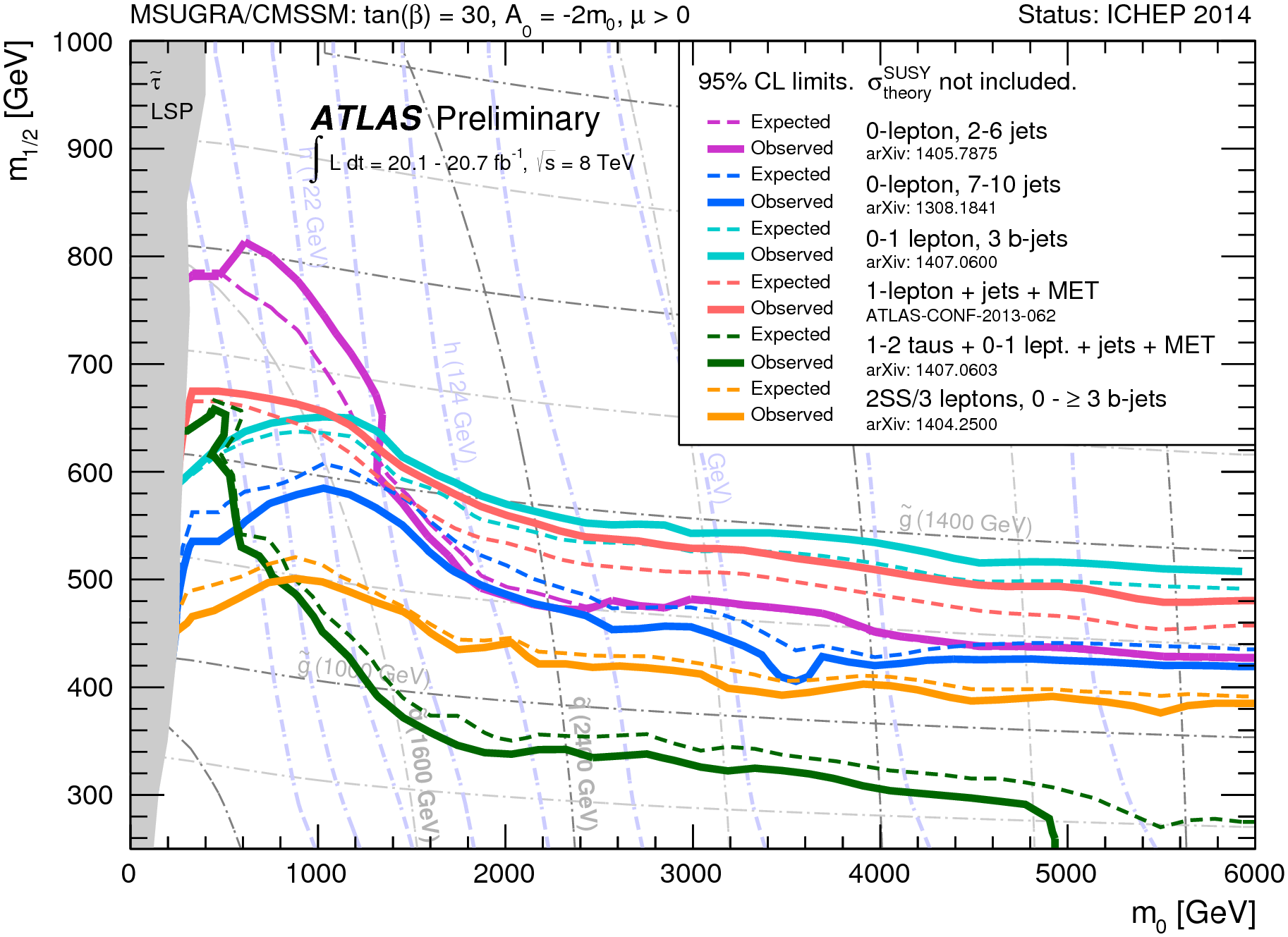}{0.95}
\caption{
Regions of the $m_0$ vs. $m_{1/2}$ plane which are excluded by various Atlas
experiment searches for gluino and squark cascade decay signatures in the mSUGRA
model with $\tan\beta =30$ and $A_0=-2m_0$. Limits are for $\sqrt{s}=8$ TeV and 20 fb$^{-1}$.
\label{fig:atlas}}
\end{figure}

A vast array of further searches have taken place: for electroweak -inos, top and bottom squarks
and sleptons in mSUGRA and in simplified models and for SUSY particle production in
a variety of different models. 
A compendium of limits can be found {\it e.g.} in Ref. \cite{pdb}.

\subsection{SUGRA gauge theories: natural or not?}

While SUGRA gauge theories are both elegant and supported indirectly by data, 
they have come under a growing body of criticism due to a perception
that they are increasingly unnatural with respect to the weak scale 
in light of recent LHC results on the Higgs mass and on lack of signal 
for sparticles (for just a few examples, 
see Ref's \cite{Shifman:2012na,Barbieri:2013vca,Giudice:2013yca,Altarelli:2014roa,Craig:2013cxa,Murayama:2014ita,Ross:2014mua,Lykken:2014bca,Dine:2015xga}).
The increasing gap between the sparticle mass scale and the weak scale is
frequently referred to as the supersymmetric Little Hierarchy Problem (LHP)\cite{barbstrum,radpq}.
To see how this comes about, we must scrutinize several measures of naturalness\cite{comp,azar_xt,dew}.
\footnote{Further investigations into naturalness in SUSY theories include Ref's 
\cite{kane,ac1,dg,ccn,ellis2,king,casas,fp,Harnik:2003rs,Nomura:2005qg,Athron:2007ry,ross,derm_kim,Allanach:2012vj,shafi,perel,Wymant:2012zp,antusch,Kribs:2013lua,hardy,Galloway:2013dma,Craig:2013fga,sug19,Fichet:2012sn,Younkin:2012ui,Kowalska:2013ica,han,Krippendorf:2012ir,Dudas:2013pja,Arvanitaki:2013yja,Fan:2014txa,Gherghetta:2014xea,Kowalska:2014hza,feng,ltr,rns,comp,deg,Martin:2013aha,Fowlie:2014xha,
Casas:2014eca}.}

But first, an important point to be made is that {\it any} quantity can look 
fine-tuned if one splits it into {\it dependent} pieces.
By re-writing an observable ${\cal O}$ as ${\cal O}+b -b$ 
and allowing $b$ to be large, 
the quantity might look fine-tuned. In this trivial example, however, combining 
dependent contributions into independent units ($b-b=0$) obviously erases the presumed source of fine-tuning.
To avoid such pitfalls, a simple fine-tuning rule has been proposed\cite{dew}:
\begin{quotation}
When evaluating fine-tuning, it is not permissible to claim fine-tuning 
of {\it dependent} quantities one against another.
\end{quotation}

\subsubsection{The electroweak measure $\Delta_{EW}$}

The electroweak measure, $\Delta_{EW}$\cite{ltr,rns}, implements the 
Dimopoulos-Susskind requirement that there be no large/unnatural cancellations
in deriving the value of $m_Z$ from the weak scale scalar potential:
\bea
\frac{m_Z^2}{2} &=& \frac{(m_{H_d}^2+\Sigma_d^d)-(m_{H_u}^2+\Sigma_u^u)\tan^2\beta}{(\tan^2\beta -1)}
-\mu^2\\
&\simeq &-m_{H_u}^2-\mu^2-\Sigma_u^u
\label{eq:mzs}
\eea
where $m_{H_u}^2$ and $m_{H_d}^2$ are the {\it weak scale} soft SUSY breaking Higgs masses, $\mu$
is the {\it supersymmetric} higgsino mass term and $\Sigma_u^u$ and $\Sigma_d^d$ contain
an assortment of loop corrections to the effective potential. The $\Delta_{EW}$ measure 
compares the largest contribution on the right-hand-side of Eq. \ref{eq:mzs} 
to the value of $m_Z^2/2$. If they are comparable, then no
unnatural fine-tunings are required to generate $m_Z=91.2$ GeV. 
The main requirement is then that
$|\mu |\sim m_Z$\cite{Chan:1997bi,Barbieri:2009ev,hgsno} (with $\mu \agt 100$ GeV to accommodate LEP2 limits 
from chargino pair production searches) 
and also that $m_{H_u}^2$ is driven radiatively to small, 
and not large, negative values~\cite{ltr,rns}.\footnote{Some recent work on theories with naturalness and
heavy higgsinos include \cite{Cohen:2015ala,Nelson:2015cea}.}
This can always happen in models where the Higgs soft terms are non-universal, such as in the 
two-extra parameter non-universal Higgs model NUHM2\cite{nuhm2}. 
Also, the top squark contributions to the radiative corrections $\Sigma_u^u(\tst_{1,2})$ 
are minimized for TeV-scale highly mixed top squarks\cite{ltr}. This latter condition  also lifts 
the Higgs mass in Eq. \ref{eq:mhiggs} to $m_h\sim 125$ GeV.
The measure $\Delta_{EW}$ is pre-programmed in the Isasugra SUSY spectrum generator\cite{isajet}.

One advantage of $\Delta_{EW}$ is that--
within the context of the MSSM-- it is 1. {\it model-independent}: if a weak scale spectrum is generated
within the pMSSM or via some high scale constrained model, one obtains exactly the
same value of naturalness. 
Other virtues of $\Delta_{EW}$ (as discussed in Ref. \cite{rns}) are
that it is: 2. the most conservative of the three measures,
3. in principle measureable, 4. unambiguous, 5. predictive, 6. falsifiable 
and 7. simple to calculate.

The principle criticism of $\Delta_{EW}$ is that-- since it involves only weak scale parameters-- 
it may not display the sensitivity of the weak scale to variations in high scale parameters. 
We will show below that the two competing measures, $\Delta_{HS}$ and $\Delta_{BG}$, if implemented
properly according to the fine-tuning rule, essentially reduce  to $\Delta_{EW}$
so that in fact $\Delta_{EW}$ portrays the entirety of electroweak naturalness.

\subsubsection{Large-log measure  $\Delta_{HS}$}

The Higgs mass fine-tuning measure, $\Delta_{HS}$, compares the radiative correction of the 
$m_{H_u}^2$ soft term, $\delta m_{H_u}^2$,  to the physical Higgs mass
\be
m_h^2\simeq \mu^2+m_{H_u}^2(\Lambda )+\delta m_{H_u}^2 .
\label{eq:mhs}
\ee
If we assume the MSSM is valid up to some high energy scale $\Lambda$ (which may be as high as $m_{GUT}$ or even $m_P$), 
then the value of $\delta m_{H_u}^2$ can be found by integrating the renormalization group equation (RGE):
\be
\frac{dm_{H_u}^2}{dt}=\frac{1}{8\pi^2}\left(-\frac{3}{5}g_1^2M_1^2-3g_2^2M_2^2+\frac{3}{10}g_1^2 S+3f_t^2 X_t\right)
\label{eq:mHu}
\ee
where $t=\ln (Q^2/Q_0^2)$,
$S=m_{H_u}^2-m_{H_d}^2+Tr\left[{\bf m}_Q^2-{\bf m}_L^2-2{\bf m}_U^2+{\bf m}_D^2+{\bf m}_E^2\right]$
and $X_t=m_{Q_3}^2+m_{U_3}^2+m_{H_u}^2+A_t^2$.
By neglecting gauge terms and $S$ ($S=0$ in models with scalar soft term universality 
but can be large in models with non-universality),
and also neglecting the $m_{H_u}^2$ contribution to $X_t$ and the fact that $f_t$ and the soft terms
evolve under $Q^2$ variation,
then a simple expression may be obtained by integrating from $m_{SUSY}$ to the cutoff $\Lambda$:
\be
\delta m_{H_u}^2 \sim -\frac{3f_t^2}{8\pi^2}(m_{Q_3}^2+m_{U_3}^2+A_t^2)\ln\left(\Lambda^2/m_{SUSY}^2 \right) .
\label{eq:DBoE}
\ee
Here, we take as usual $m_{SUSY}^2 \simeq m_{\tst_1}m_{\tst_2}$.
By requiring\cite{kn,papucci,brust,Evans:2013jna} 
\be
\Delta_{HS}\sim \delta m_{H_u}^2/(m_h^2/2)\alt 10
\ee 
then one expects the three third generation squark masses $m_{\tst_{1,2},\tb_1}\alt 600$ GeV.
Using the $\Delta_{HS}$ measure of fine-tuning 
along with $m_h\simeq 125$ GeV, one finds some popular SUSY 
models to be electroweak fine-tuned to 0.1\%\cite{comp}.

Two pitfalls occur within this approach, which are {\it different} from the case of the SM.
\bi
\item The first is that $m_{H_u}^2(\Lambda )$ and $\delta m_{H_u}^2$ are {\it not} independent:
the value of $m_{H_u}^2$ feeds directly into evaluation of $\delta m_{H_u}^2$ via the $X_t$ term.
It also feeds indirectly into $\delta m_{H_u}^2$ by contributing to the evolution of the
$m_{Q_3}^2$ and $m_{U_3}^2$ terms. 
This can be seen in Fig. \ref{fig:mHu2} where we plot $\delta m_{H_u}^2$ as a function of $m_{H_u}^2(m_{GUT} )$
for a particular choice of model parameters.
We see that as $m_{H_u}^2(m_{GUT})$ increases, then there is an increasingly {\it negative}
cancelling correction $\delta m_{H_u}^2$. 
Thus, this measure {\it fails} the fine-tuning rule\cite{dew}.
\ei
\begin{figure}[tbp]
\postscript{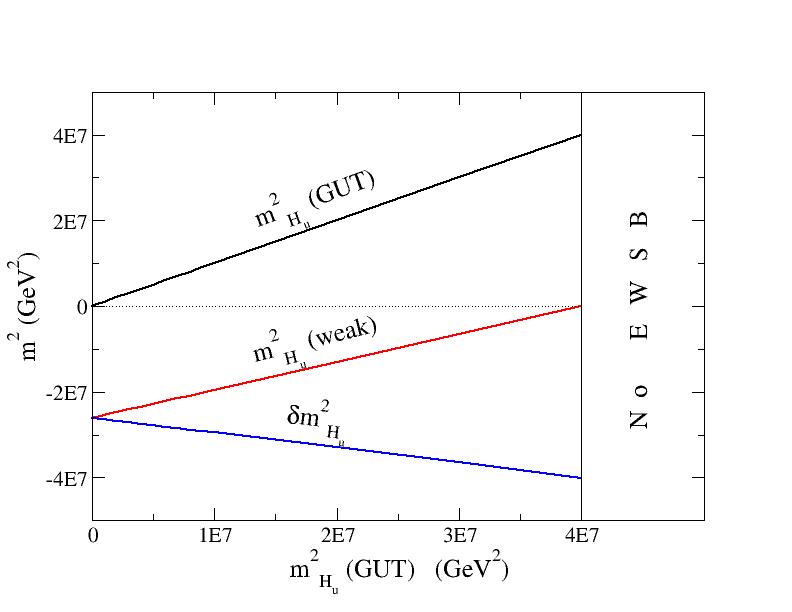}{1.05}
\caption{
Values of $\delta m_{H_u}^2$, $m_{H_u}^2 (m_{GUT})$ and $m_{H_u}^2 (m_{weak})$
vs. $m_{H_u}^2 (m_{GUT})$ for a model with $m_0=5$ TeV, $m_{1/2}=700$ GeV and
$A_0=-8$ TeV and $\tan\beta =10$. Here, $m_{H_u}^2(weak)=m_{H_u}^2(m_{GUT})+\delta m_{H_u}^2$.
\label{fig:mHu2}}
\end{figure}
\bi
\item A second issue with $\Delta_{HS}$ is that-- whereas $SU(2)_L\times U(1)_Y$ gauge symmetry can be broken at tree
level in the SM-- in the SUGRA case, where SUSY is broken in a hidden sector via the superHiggs mechanism, 
then $m_{H_u}^2\sim m_{3/2}^2>0$. Thus, for SUGRA models, electroweak symmetry is not even broken until
one includes radiative corrections. For SUSY models valid up to some high scale $\Lambda\gg m_{weak}$,
the large log in Eq.~\ref{eq:DBoE} is exactly what is required to break EW symmetry in the first place,
radiatively driving $m_{H_u}^2$ to negative values\cite{rewsb}.
\ei

A simple fix for $\Delta_{HS}$ is to {\it combine the dependent terms} into a single quantity. 
Under such a regrouping\cite{ltr,rns}, then 
\be
m_h^2|_{phys}=\mu^2+\left(m_{H_u}^2(\Lambda )+\delta m_{H_u}^2 \right)
\label{eq:mh}
\ee
where now $\mu^2$ and $\left(m_{H_u}^2(\Lambda )+\delta m_{H_u}^2 \right)$ are each independent so each 
should be comparable to $m_h^2$ in order to avoid fine-tuning. 
The large log is still present in $(m_{H_u}^2(\Lambda )+\delta m_{H_u}^2)$, but now cancellations can occur
between the boundary condition and the radiative correction.

It is sometimes claimed that under such a regrouping, then the SM Higgs mass would also not be fine-tuned. 
But here we see that, in the MSSM case-- since the $m_{H_u}^2$ and $\delta m_{H_u}^2$ terms are dependent-- 
the situation is different from the SM and one must combine dependent terms together.
The regrouping of contributions to $m_h^2$ leads back to the $\Delta_{EW}$ measure since 
now $(m_{H_u}^2(\Lambda )+\delta m_{H_u}^2) = m_{H_u}^2(weak)$. 
Indeed, we see from Fig. \ref{fig:mHu2} that for large enough $m_{H_u}^2( m_{GUT})$, 
then there can be a large cancellation so that the weak scale value of
$|m_{H_u}^2|$ is indeed comparable to $m_Z^2$. 
This is the case of radiatively-driven naturalness.

\subsubsection{Sensitivity to high scale parameters:  $\Delta_{BG}$}

The more traditional measure, $\Delta_{BG}$, was proposed by Ellis {\it et al.}\cite{Ellis:1986yg} 
and later investigated more thoroughly by Barbieri and Giudice\cite{bg}. 
The starting point is to express $m_Z^2$ in terms of weak scale SUSY parameters
as in Eq. \ref{eq:mzs}:
\be
m_Z^2 \simeq -2m_{H_u}^2-2\mu^2 ,
\label{eq:mZsapprox}
\ee
where the partial equality obtains for moderate-to-large $\tan\beta$ values and where we assume for
now the radiative corrections are small.
To evaluate $\Delta_{BG}$, one needs to know the explicit dependence of 
the weak scale values of $m_{H_u}^2$ and $\mu^2$ on the
fundamental parameters. 
Semi-analytic solutions to the one-loop renormalization group equations
for $m_{H_u}^2$ and $\mu^2$ can be found for instance in Ref's \cite{munoz}.
For the case of $\tan\beta =10$, then\cite{abe,martin,feng}
\begin{eqnarray}
&&\mbox{\hspace{-.2cm}}m_Z^2 = -2.18\mu^2 + 3.84 M_3^2+0.32M_3M_2 \nonumber \\ 
&&\mbox{\hspace{-.3cm}}+0.047 M_1M_3 -0.42 M_2^2 +0.011 M_2M_1-0.012M_1^2 \nonumber \\
&&\mbox{\hspace{-.3cm}}-0.65 M_3A_t -0.15 M_2A_t-0.025M_1 A_t+0.22A_t^2 \nonumber \\
&&\mbox{\hspace{-.3cm}}+0.004 M_3A_b -1.27 m_{H_u}^2 -0.053 m_{H_d}^2 \nonumber \\
&&\mbox{\hspace{-.3cm}}+.73 m_{Q_3}^2+.57 m_{U_3}^2+.049 m_{D_3}^2-.052 m_{L_3}^2+.053 m_{E_3}^2\nonumber \\
&&
\mbox{\hspace{-.3cm}}+.051 m_{Q_2}^2-.11 m_{U_2}^2+.051 m_{D_2}^2-.052 m_{L_2}^2+.053 m_{E_2}^2\nonumber \\
&&\mbox{\hspace{-.3cm}}
+.051 m_{Q_1}^2-.11 m_{U_1}^2+.051 m_{D_1}^2-.052 m_{L_1}^2+.053 m_{E_1}^2,\mbox{\hspace{-.3cm}}\nonumber\\
&&
\label{eq:mzslong}
\end{eqnarray}
where the parameters on the right-hand-side are understood as 
evaluated at the GUT scale.
(For different values of $\tan\beta$, then somewhat different co-efficients are obtained.)

Then, the proposal is that the variation in $m_Z^2$ with respect to
parameter variation be small:
\be
\Delta_{BG}\equiv max_i\left[ c_i\right]\ \ {\rm where}\ \ 
c_i=\left|\frac{\partial\ln m_Z^2}{\partial\ln p_i}\right|
=\left|\frac{p_i}{m_Z^2}\frac{\partial m_Z^2}{\partial p_i}\right|
\label{eq:DBG}
\ee
where the $p_i$ constitute the fundamental parameters of the model.
Thus, $\Delta_{BG}$ measures the fractional change in $m_Z^2$ due to fractional variation in
high scale parameters $p_i$.
The $c_i$ are known as {\it sensitivity coefficients}\cite{bg}. 

The requirement of low $\Delta_{BG}$ is then equivalent to the requirement of no 
large cancellations on the right-hand-side of Eq. \ref{eq:mzslong} since (for linear terms) 
the logarithmic derivative just picks off coefficients of the relevant parameter. For instance, 
$c_{m_{Q_3}^2}=0.73\cdot (m_{Q_3}^2/m_Z^2)$. If one allows $m_{Q_3}\sim 3$ TeV (in accord with 
requirements from the measured value of $m_h$), then one obtains $c_{m_{Q_3}^2}\sim 800$
and so $\Delta_{BG}\ge 800$. In this case, SUSY would be electroweak fine-tuned to about 0.1\%. 
If instead one insists that $m_0$ is the fundamental parameter with $m_{Q_3}=m_{U_3}=m_{H_u}\equiv m_0$, 
as in models with scalar mass universality, 
then the various scalar mass contributions to $m_Z^2$ largely {\it cancel} and $c_{m_0^2}\sim -0.017 m_0^2/m_Z^2$: 
the contribution to $\Delta_{BG}$ from scalars drops by a factor $\sim 50$\cite{feng}. 

The above example illustrates the extreme model-dependence of $\Delta_{BG}$ for multi-parameter SUSY models. 
The value of $\Delta_{BG}$ can change radically from theory to theory even if those theories 
generate exactly the same weak scale sparticle mass spectrum. The model dependence of $\Delta_{BG}$
arises due to a violation of the Fine-tuning Rule: one must combine dependent terms into independent quantities
before evaluating EW fine-tuning.

\subsubsection{$\Delta_{BG}$ applied to SUGRA gauge theories}

In Ref. \cite{comp}, it was argued that: in an ultimate theory (UTH), where all soft parameters are correlated, 
then $\Delta_{BG}$ should be a reliable measure of naturalness. In fact, SUGRA gauge theories with 
hidden sector SUSY breaking fulfill this requirement. 
The amazing thing is that we do not need to know the precise hidden sector in order to properly evaluate $\Delta_{BG}$. 

In supergravity gauge theories with hidden sector SUSY breaking via the superHiggs mechanism,
where the hidden sector is fully specified,  
the gravitino gains a mass $m_{3/2}$ but then in addition all soft SUSY breaking terms are generated as multiples 
of the gravitino mass $m_{3/2}$. 
Thus, we can write each soft term as
\bea
m_{H_u}^2&=&a_{H_u}\cdot m_{3/2}^2,\label{eq:1} \\
m_{Q_3}^2&=&a_{Q_3}\cdot m_{3/2}^2,\\
A_t&=&a_{A_t}\cdot m_{3/2},\\
M_i&=&a_i\cdot m_{3/2},\\
& & \cdots \label{eq:5} .
\eea
For any fully specified hidden sector, the various $a_i$ are calculable.
For example, in string theory with dilaton-dominated SUSY breaking\cite{kl,Brignole:1993dj}, 
we expect $m_0^2=m_{3/2}^2$ with $m_{1/2}=-A_0=\sqrt{3}m_{3/2}$. 
Alternatively, acknowledging our lack of knowledge of hidden sector dynamics, 
we may {\it parametrize our ignorance} by leaving the $a_i$ as free parameters. 
By using several adjustable parameters, we cast a wide net which encompasses a large range of hidden sector SUSY breaking possibilities. 
But this doesn't mean that each SSB parameter is expected to be independent of the 
others.
It just means we do not know how SUSY breaking occurs, and how the soft terms are correlated:
it is important not to confuse parameters, which ought to be related to one another 
in any sensible theory of SUSY breaking, with independently adjustable soft SUSY breaking terms.

Now, plugging the soft terms Eq's \ref{eq:1}-\ref{eq:5} into Eq. \ref{eq:mzslong}, one arrives at the simpler expression
\be
m_Z^2=-2.18\mu^2 +a\cdot m_{3/2}^2 .
\label{eq:mzssugra}
\ee
The value of $a$ is just some number which is the sum of all the coefficients of the terms $\propto m_{3/2}^2$.
For now, we assume $\mu$ is independent of $m_{3/2}$ as will be discussed below.

Using Eq. \ref{eq:mzssugra}, we can compute the sensitivity coefficients in the theory 
where the soft terms are properly correlated:\footnote{In mAMSB, the soft terms are 
also written as multiples of $m_{3/2}$ or $m_{3/2}^2$. In mGMSB, the soft terms are written as multiples
of messenger scale $\Lambda_m$. The argument proceeds in an identical fashion in these cases.}
\bea
c_{m_{3/2}^2} &=& |a\cdot (m_{3/2}^2/m_Z^2)|\ \ {\rm and}\label{eq:A} \\
c_{\mu^2} &=& |-2.18 (\mu^2/m_Z^2 )|.\label{eq:B}
\eea
For $\Delta_{BG}$ to be $\sim 1-10$ (natural SUSY with low fine-tuning), then Eq. \ref{eq:B} implies
\bi
\item $\mu^2 \sim m_Z^2$  .
\ei
Also, Eq. \ref{eq:A} implies 
\bi
\item $a\cdot m_{3/2}^2\sim m_Z^2$.
\ei 
The first of these conditions implies light higgsinos with mass $\sim 100-200$ GeV, the closer to $m_Z$ 
the better. The second condition can be satisfied if $m_{3/2}\sim m_Z$\cite{bg} 
(which now seems highly unlikely
due to a lack of LHC8 SUSY signal\footnote{For instance, in simple SUGRA models, the scalar masses $m_0=m_{3/2}$. 
Since LHC requires rather high $m_0$, then we would also expect rather large $m_{3/2}$.} 
and the rather large value of $m_h$) {\it or} if $a$ is quite small:
in this latter case, the SSB terms conspire such that there are large cancellations amongst the various
coefficients of $m_{3/2}^2$ in Eq. \ref{eq:mzslong}: 
this is what is called radiatively-driven natural SUSY\cite{ltr,rns} since in this case a large high scale 
value of $m_{H_u}^2$ can be driven radiatively to small values $\sim -m_Z^2$ at the weak scale.

Furthermore, we can equate the value of $m_Z^2$ in terms of weak scale parameters with the value 
of $m_Z^2$ in terms of GUT scale parameters:
\bea
m_Z^2&\simeq &-2\mu^2(weak)-2m_{H_u}^2(weak) \nonumber\\
&\simeq &-2.18\mu^2(GUT)+a\cdot m_{3/2}^2 .
\eea
Since $\mu$ hardly evolves under RG running (the factor 2.18 is nearly 2), then we have the
BG condition for low fine-tuning as 
\be
-2m_{H_u}^2(weak) \sim a\cdot m_{3/2}^2\sim m_Z^2 ,
\ee
{\it i.e.} that the value of $m_{H_u}^2$ must be driven to 
{\it small} negative values $\sim - m_Z^2$ at the weak scale. 
These are exactly the conditions required by the model-independent EWFT
measure $\Delta_{EW}$: {\it i.e.} we have
\be
\lim_{n_{SSB}\to 1} \Delta_{BG}\to \Delta_{EW}
\ee
where $n_{SSB}$ is the number of {\it independent} soft SUSY breaking terms.
In this sense, a low value of $\Delta_{EW}$ reflects not only low weak scale
fine-tuning, but also low high scale fine-tuning!
Of course, this approach also reconciles the Higgs mass fine-tuning measure $\Delta_{HS}$ (with
appropriately regrouped independent terms) with the $\Delta_{BG}$ measure 
(when applied to models with a single independent soft breaking term such as $m_{3/2}$).


%
%
%

\subsubsection{A worked example: BG and EW fine-tuning in a model 
with a Polonyi-type hidden sector}

As a concrete example, we evaluate $\Delta_{BG}$ in a model with
$m_{\tq}=m_{\tell}=m_{H_{u,d}}=1$ TeV, $A_0=1268$ GeV and $\tan\beta =10$ and with
$m_{1/2}=m_{\tq}/3$. In the MSSM, the largest sensitivity co-efficient
comes from the $m_{H_u}^2$ term in Eq. \ref{eq:mzslong} yielding $\Delta_{BG}=\Delta_{H_u}=1.27 m_{H_u}^2(m_{GUT})/m_Z^2=153$.
If we notice the squark and slepton masses are universal, we may instead evaluate in the NUHM2 model
where the combined 3rd generation terms in Eq. \ref{eq:mzslong} give the largest contribution: 
$\Delta_{BG}=\Delta_{m_0(3)}=1.35 m_0^2/m_Z^2=162$. 
If we further notice that the Higgs and matter scalar soft terms are degenerate at $m_{GUT}$ 
($m_{H_u}^2=m_{H_d}^2=m_0^2$), we may
instead evaluate in the mSUGRA/CMSSM model. In this case, the combined third generation and Higgs 
contributions to Eq. \ref{eq:mzslong} largely cancel so that the largest soft term contribution
comes from $\Delta_{m_{1/2}}=m_{1/2}(7.572 m_{1/2}-0.821 A_0)/(2m_Z^2)=29.6$. 
However, $\mu^2$ should really be regarded an input parameter where $\mu (gut)=445$ GeV.
In this case, $\Delta_{BG}=\Delta_{\mu^2}=2.18\mu^2(gut)/m_Z^2=51.9$.
If we make a final realization that the soft terms are exactly those
of the Polonyi model Eq. \ref{eq:polonyi}-- all computed as multiples of $m_{3/2}$ 
with $m_{1/2}=m_{3/2}/3$-- 
then $\Delta_{BG}=\Delta_{m_{3/2}}=0.44 m_{3/2}^2/m_Z^2=52.9$. 
These values are displayed in the histograms of Fig. \ref{fig:polonyi}.
We also compare against the value of $\Delta_{EW}\simeq \mu^2(weak)/(m_Z^2/2)=51.9$
since $\mu (weak) =465$ GeV in order to enforce $m_Z=91.2$ GeV.
Thus, we see that $\Delta_{EW}$ is a good approximation to $\Delta_{BG}$ 
when evaluated using Eq. \ref{eq:mzssugra} where $a$ turns out to be $0.44$.
\begin{figure}[tbp]
\postscript{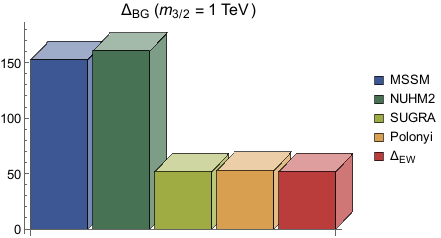}{0.95}
\caption{
Values of $\Delta_{BG}$ computed in a model with 
$m_{\tq}=m_{\tell}=m_{H_{u,d}}=1$ TeV, $A_0=1268$ GeV and $\tan\beta =10$ with
$m_{1/2}=m_{\tq}/3$.
We show $\Delta_{BG}$ as computed for the MSSM model, 
the NUHM2 model, the mSUGRA model and the Polonyi
model with $m_{3/2}=1$ TeV. 
We also show the value of $\Delta_{EW}$.
\label{fig:polonyi}}
\end{figure}

\subsection{Radiatively-driven naturalness}

We have seen that, when applied appropriately, the three measures
of SUSY weak scale naturalness are in accord:
\be
\Delta_{EW}\simeq \Delta_{HS}\simeq \Delta_{BG} .
\ee
Thus, in the following discussion we will use the EW measure due to the ease
of including radiative corrections: the 43 terms of $\Sigma_u^u$ and $\Sigma_d^d$
which are listed in the Appendix of Ref. \cite{rns}.
The requirements for natural SUSY are then plain to see:
\bi
\item the soft term $m_{H_u}^2$ is driven radiatively to small
negative values $\sim -m_Z^2$ at the weak scale,
\item the $\mu$ parameter independently is of magnitude $\sim m_Z$,
the closer to $m_Z$ the better, and
\item the radiative corrections $|\Sigma_u^u |$ should be not much
larger than $m_Z^2$. The largest contribution to $\Sigma_u^u$ comes almost
always from the stop sector. Using the exact one-loop radiative corrections, 
the large $A_t$ trilinear soft term suppresses both the terms 
$\Sigma_u^u(\tst_{1,2})$ whilst lifting $m_h$ to $\sim 125$ GeV\cite{ltr}. 
Since under RG running the gluino soft term $M_3$ lifts the stop masses, a 
limit on the contribution $\Sigma_u^u (\tst_{1,2})$ also provides a 
(two-loop) upper bound on $m_{\tg}$.
\ei
The term $m_{H_d}^2$ is suppressed by $\tan^2\beta$ in Eq. \ref{eq:mzs}
and so can be much larger: in the multi-TeV range without violating naturalness.
Since the heavy SUSY Higgs masses $m_{A,H,H^\pm}\sim m_{H_d}$, then these 
could all live in the TeV range, perhaps beyond the reach of 
LHC13\cite{rnshiggs}.

A scan over NUHM2 parameter space yields the plot of $m_{\tg}$ vs. $\Delta_{EW}$
in Fig. \ref{fig:mgl}. Here, we see that for $\Delta_{EW}<30$, the 
upper bound on $m_{\tg}$ extends to $\sim 4 $ TeV, well beyond the ultimate reach of LHC. If instead we require $\Delta_{EW}\alt 10$, then $m_{\tg}\alt 2$ TeV, and
should be accessible to LHC searches. 
In Fig. \ref{fig:mt1}, we show the value of $m_{\tst_1}$ vs. $\Delta_{EW}$.
Here, we see that light stop masses can exist in the  1-2 TeV range 
while maintaining naturalness. 
In Fig. \ref{fig:mw1}, we show the mass of the lightest charged higgsino
$m_{\tw_1}$ vs. $\Delta_{EW}$. For $\Delta_{EW}\alt 30$, then $m_{\tw_1}\alt 300$ GeV
so that a linear $e^+e^-$ collider operating with $\sqrt{s}\sim 600$ GeV
will probe the entire space with modest values of $\Delta_{EW}$.
\begin{figure}[tbp]
\postscript{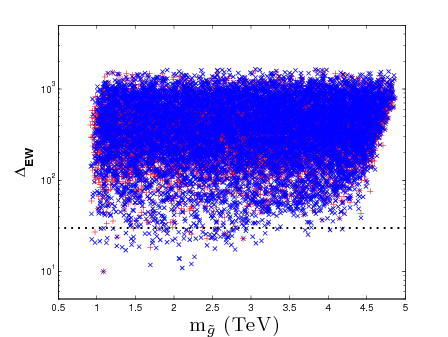}{1.05}
\caption{
Plot of $m_{\tg}$ vs. $\Delta_{EW}$ from a scan over NUHM2 parameter space
(Ref. \cite{rns}). 
Blue points repreent a focussed scan at low $\mu$ while red points represent a scan over a broader range of $\mu$.
\label{fig:mgl}}
\end{figure}
\begin{figure}[tbp]
\postscript{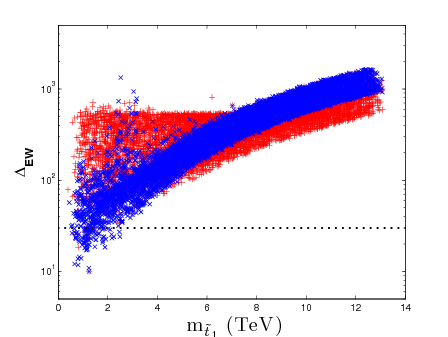}{1.05}
\caption{
Plot of $m_{\tst_1}$ vs. $\Delta_{EW}$ from a scan over NUHM2 parameter space
(Ref. \cite{rns}).
Blue points repreent a focussed scan at low $\mu$ while red points represent a scan over a broader range of $\mu$.
\label{fig:mt1}}
\end{figure}
\begin{figure}[tbp]
\postscript{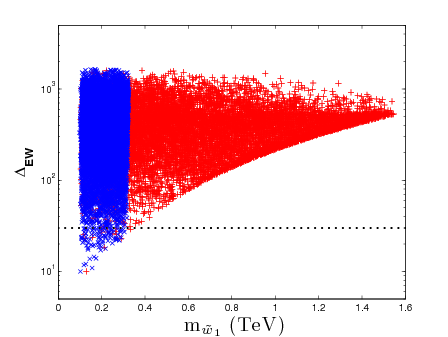}{1.05}
\caption{
Plot of $m_{\tw_1}$ vs. $\Delta_{EW}$ from a scan over NUHM2 parameter space
(Ref. \cite{rns}).
Blue points repreent a focussed scan at low $\mu$ while red points represent a scan over a broader range of $\mu$.
\label{fig:mw1}}
\end{figure}

A typical sparticle mass spectrum with radiatively-driven naturalness
is shown in Fig. \ref{fig:spec}.
\begin{figure}[tbp]
\postscript{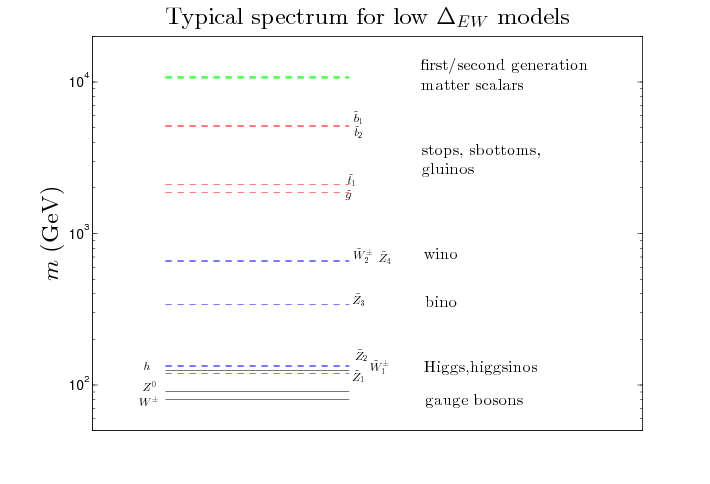}{1.05}
\caption{
Typical sparticle mass spectrum from SUSY models with low $\Delta_{EW}$,
{\it i.e.} radiatively-driven naturalness.
\label{fig:spec}}
\end{figure}

\subsection{QCD naturalness, Peccei-Quinn symmetry and 
the $\mu$ problem}

While on the topic of naturalness, we should include discussion of naturalness in the QCD sector.
In the early days of QCD, it was a mystery why the two-light-quark chiral symmetry $U(2)_L\times U(2)_R$
gave rise to three and not four light pions\cite{U1}. 
The mystery was resolved by 't Hooft's discovery of the QCD theta vacuum which didn't respect 
the $U(1)_A$ symmetry\cite{tHooft}. 
As a consequence of the theta vacuum, one expects the presence of a term 
\be
{\cal L}\ni \frac{\bar{\theta}}{32\pi^2}F_{A\mu\nu}\tilde{F}_A^{\mu\nu}
\ee 
in the QCD Lagrangian (where $\bar{\theta}=\theta+arg(det({\cal M}))$ and ${\cal M}$ 
is the quark mass matrix). Measurements of the neutron EDM constrain $\bar{\theta}\alt 10^{-10}$ 
leading to an enormous fine-tuning in $\bar{\theta}$: the so-called strong CP problem.

The strong CP problem is elegantly solved via the PQWW\cite{pqww} introduction of PQ symmetry 
and the concomitant (invisible\cite{ksvz,dfsz}) axion: 
the offending term can dynamically settle to zero.
The axion is a valid dark matter candidate in its own right\cite{axdm}.

Introducing the axion in a SUSY context solves the strong CP problem but also offers an 
elegant solution to the SUSY $\mu$ problem. The most parsimonius implementation of the strong CP solution
involves introducing a single MSSM singlet superfield $S$ carrying PQ charge $Q_{PQ}=-1$ while the
Higgs fields both carry $Q_{PQ}=+1$. The usual mu term is forbidden, but then we have a 
superpotential\cite{susydfsz}
\be
W_{DFSZ}\ni \lambda\frac{S^2}{M_P}H_uH_d .
\ee
If PQ symmetry is broken and $S$ receives a VEV $\langle S\rangle\sim f_a$, then a weak scale
mu term
\be
\mu\sim \lambda f_a^2/M_P
\ee
is induced which gives $\mu\sim m_Z$ for $f_a\sim 10^{10}$ GeV. While Kim-Nilles sought to relate
the PQ breaking scale $f_a$ to the hidden sector mass scale $m_{hidden}$\cite{KN}, we see now that the
Little Hierarchy 
\be
\mu\sim m_Z\ll m_{3/2}\sim {\rm multi-TeV} 
\ee
could emerge due to a mis-match between PQ breaking scale and hidden sector mass scale $f_a\ll m_{hidden}$.

In fact, an elegant model which exhibits this behavior was put forth by 
Murayama, Sakai and Yanagida (MSY)\cite{msy}. In the MSY model, PQ symmetry is broken
radiatively by driving one of the PQ scalars $X$ to negative mass-squared values in much the same way
that electroweak symmetry is broken by radiative corrections driving $m_{H_u}^2$ negative.
Starting with multi-TeV scalar masses, the radiatively-broken PQ symmetry induces 
a SUSY $\mu $ term $\sim 100$ GeV\cite{radpq} while at  the same time generating 
intermediate scale Majorana masses for right-hand neutrinos: see Fig. \ref{fig:run_m32}.
In models such as MSY, the Little Hierarchy $\mu\ll m_{3/2}$ is no problem at all 
but is instead just a reflection of the mis-match between PQ and hidden sector mass scales.
\begin{figure}[tbp]
\postscript{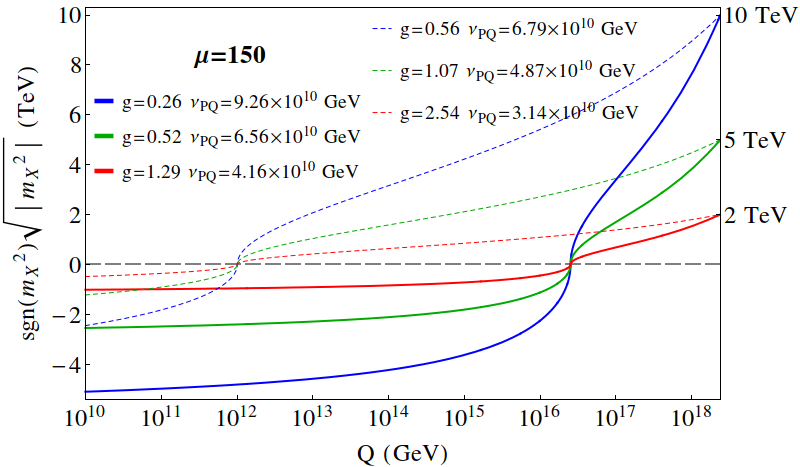}{1.0}
\caption{Plot of the running values of $m_X^2$ versus $Q$ for various 
values of $m_{3/2}$ and $h=2$ (dashed) and $h=4$ (solid).
Here, $g$ and $h$ are couplings from the MSY model Lagrangian\cite{msy,radpq}.
\label{fig:run_m32}}
\end{figure}

\section{Implications}

\subsection{LHC searches for SUSY with radiatively-driven naturalness}

SUSY models with radiatively-driven naturalness have been examined in the case of 
gaugino mass unification (where the LSP is a higgsino-like neutralino) 
and in lesser detail for the case of non-unified gaugino masses (where the LSP
could be either bino-like or wino-like while preserving naturalness)\cite{nugm}.
Here, we briefly summarize prospects for the more motivated case with gaugino mass unification.

For sparticle searches at LHC13, the best prospects for the next couple years will be in searches
for gluino and squark pair production. While squark masses can range into the 10-20 TeV range
while not compromising naturalness\footnote{Such heavy matter scalars provide a decoupling solution\cite{dine}
to the SUSY flavor and $CP$ problems and are favored by a heavy gravitino solution to the gravitino problem.
However, such heavy squarks/sleptons can lead to large loop-level contributions to $m_Z$ unless
certain GUT relations amongst masses are obeyed\cite{deg}.}, gluino masses are required to be below about 2 TeV 
for $\Delta_{EW}\alt 10$ and less than about 5 TeV for $\Delta_{EW}\alt 30$\cite{rns}. Thus, a lucrative portion
of RNS parameter space will be accessible via gluino pair searches at LHC13. 
In searching for $pp\to \tg\tg X$ production, in RNS models the dominant gluino decay
is to third generation quarks: $\tg\to t\tst_1$ (followed by $\tst\to b\tw_i$) 
if kinematically allowed or to three-body modes $t\bar{t}\tz_i$ or $t b\tw_i$ 
if two-body modes are closed\cite{gltotop}.
These decays will yield the usual multi-jet $+$ multi-isolated-lepton $+\eslt$ gluino cascade decay
events albeit ones that are rich in identifiable $b$-jets\cite{btw2,bcpt}. 
The mass edge at $m(\ell^+\ell^-)<m_{\tz_2}-m_{\tz_1}$ arising from $\tz_2\to\tz_1\ell\bar{\ell}$ decay\cite{bcpt} 
may be apparent in cascade decay events containing OS/SF dileptons.
The LHC reach for RNS is shown in Table \ref{tab:reach} (for $\sqrt{s}=14$ TeV) in terms of $m_{\tg}$ where
squarks are assumed very heavy\cite{lhc}. The LHC reach for gluino pair production cascade decay signatures
extends to $m_{\tg}\sim 1.9$ TeV for 1000 fb$^{-1}$ of integrated luminosity.
\begin{table}
\begin{center}
\begin{tabular}{|l|r|r|r|r|}
\hline
 Int. lum. (fb$^{-1}$) & $\tg\tg$ &  SSdB & $WZ\to 3\ell$ &$4\ell$ \\
\hline
\hline
10   & 1.4 &  --  & -- & --\\
100  & 1.6 &  1.6 & -- & $\sim 1.2$\\
300  & 1.7 &  2.1 & 1.4& $\gtrsim 1.4$ \\
1000 & 1.9 &  2.4 & 1.6& $\gtrsim 1.6$ \\
\hline
\end{tabular}
\caption{Reach of LHC14 for SUSY in terms of gluino mass, $m_{\tg}$ (TeV),  
assuming various integrated luminosity values along the RNS model line.
We present each search channel considered in this paper except soft
$3\ell$.
\label{tab:reach}}
\end{center}
\end{table}

Since the higgsino states $\tw_1^\pm$ and $\tz_{1,2}$ are so light in RNS, they tend to provide the
dominant SUSY production cross section. However, the heavier higgsino states decay via three-body mode
to lighter higgsino states: $\tw_1\to f\bar{f}'\tz_1$ and $\tz_2\to\tz_1 f\bar{f}$. 
Since the inter-higgsino mass gap is so small-- typically just 10-20 GeV-- 
there is very little visible energy release as
most of the energy goes into making up the LSP mass $m_{\tz_1}$ which serves as (a portion of) 
the dark matter. Thus, the higgsino pair production reactions seem very difficult to see at
LHC above SM processes. It is possible that making use of initial state jet radiation may help
marginally in extracting a signal for light higgsino pair production\cite{chan,monojet,kribs,dilep}.

For SUSY models with light higgsinos, a very distinctive, and ultimately more powerful,
search channel emerges: that of same-sign diboson production (SSdB)\cite{lhcltr}
as shown in Fig. \ref{fig:diagram}.
In RNS models with gaugino mass unification, the $\tw_2$ and $\tz_4$ are wino-like
and tend to provide the largest {\it visible} SUSY cross section over the 
expected range of $m_{\tg}$. This is simply because $\sigma (\tg\tg )$ is rapidly decreasing
with increasing $m_{\tg}$ and so pair production of the lighter wino-pairs $pp\to\tw_2^\pm\tz_4$
wins out. The dominant wino decay modes include $\tw_2^\pm \to\tz_{1,2} W^\pm$ and $\tz_4\to \tw_1^\pm W^\mp$.
As mentioned above, the higgsino states $\tw_1$ and $\tz_{1,2}$ yield only soft decay products and 
are quasi-invisible. The final state then consists of same-sign or opposite-sign dibosons $+\eslt$.
While the OS diboson signal is expected to be buried under a prodigious SM $W^+W^-$ background, 
the background for the same-sign diboson topology is very low. 
A detailed signal/background study in Ref's \cite{lhcltr,lhc} find the SSdB channel to 
ultimately give the best reach of LHC13 for SUSY. In Table \ref{tab:reach}, for 1000 fb$^{-1}$
the LHC14 reach via the SSdB channel extends to $m_{\tg}\sim 2.4$ TeV (compared against 1.9 TeV for
the reach via $\tg\tg$ cascade decays). While the SSdB channel gives the maximal LHC reach for RNS, it is 
also important to note that this channel is distinctive to models with light higgsinos 
and would provide strong confirmation for natural SUSY.
\begin{figure}[tbp]
\postscript{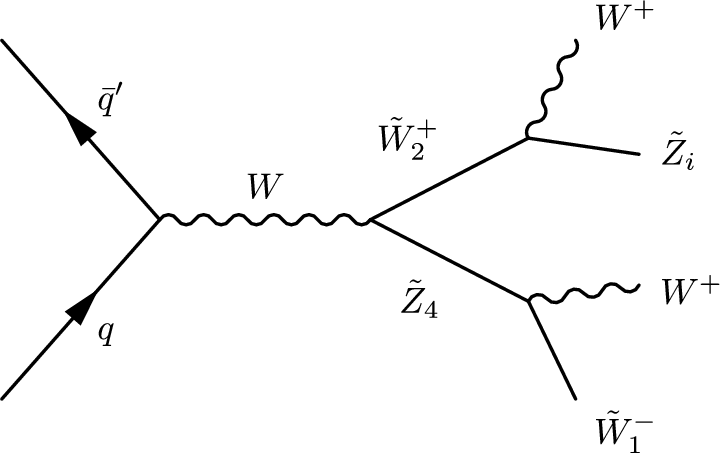}{0.9}
\caption{Diagram depicting same-sign diboson production at LHC in SUSY models 
with light higgsinos.
\label{fig:diagram}}
\end{figure}

\subsection{The ILC: a higgsino factory}

The smoking-gun signature of SUSY with radiatively-driven naturalness is the presence of
four light higgsino states $\tz_{1,2}$ and $\tw_1^\pm$ with mass $\sim |\mu |$ and where
$|\mu |\sim 100-200$ GeV. Thus, these states should be accessible to a linear $e^+e^-$ collider
operating with $\sqrt{s}>2m(higgsino)$. While the 10-20 GeV inter-higgsino mass gaps
are problematic at LHC, they should be easily visible in the clean environment of an
$e^+e^-$ collider. 

Fig. \ref{fig:ilc_xsec} shows various RNS SUSY cross sections versus $\sqrt{s}$ at the ILC.
The important point is that while one expects ILC to be constructed as a Higgs factory ($e^+e^-\to Zh$), 
it stands an excellent chance to emerge as a SUSY discovery machine and a {\it higgsino factory}!
(The limited beam energy ($\sqrt{s}\sim 350$ GeV) 
of a machine like TLEP may or may not be sufficient to produce the 
required light higgsino pairs.) From Fig. \ref{fig:ilc_xsec}, we see that the dominant
higgsino pair production reactions would consist of $e^+e^-\to\tw_1^+\tw_1^-$ and $\tz_1\tz_2$. 
Detailed studies of signal and background\cite{ilc} find that the higgsino pair production reactions
should be straightforward to extract from SM background including $\gamma\gamma$-initiated events.
Further, making use of the $\tw_1\to q\bar{q}'\tz_1$ and $\tw_1\to\ell\bar{\nu}\tz_1$ events, 
the $\tw_1$ and $\tz_1$ masses can be extracted. Also, the $\tz_2$ and $\tz_1$ masses can be extracted
from $\tz_1\tz_2$ production followed by $\tz_2\to\ell^+\ell^-\tz_1$ decay. The higgsino-like nature of the
particles is easily extracted using event kinematics and beam polarization.
As $\sqrt{s}$ is increased, further SUSY pair production reactions should successively be accessed.
\begin{figure}[tbp]
\postscript{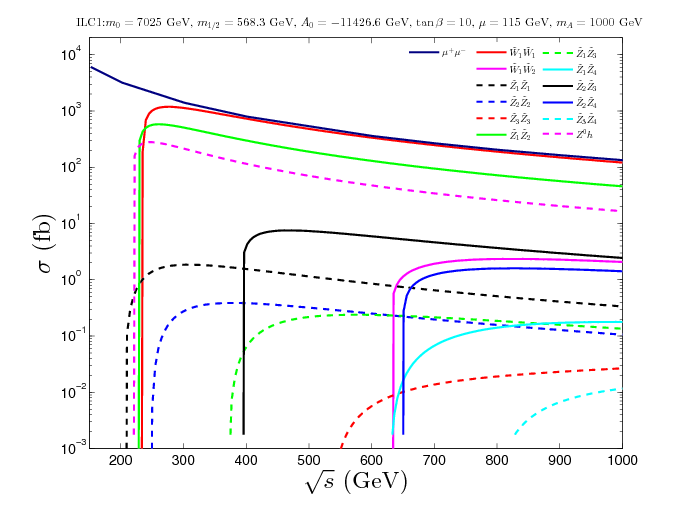}{1.0}
\caption{
{Sparticle production cross sections vs. $\sqrt{s}$ for
  unpolarized beams at the ILC $e^+e^-$ collider for an RNS benchmark point
  listed in Ref. \cite{ilc}.}
\label{fig:ilc_xsec}}
\end{figure}

\subsection{Dark matter: an axion/WIMP admixture?}

As mentioned above, to allow for both electroweak and QCD naturalness, one needs
a model including both axions and SUSY. In such a case, the axion field is promoted to a superfield
which contains a spin-0 $R$-parity even saxion $s$ and a spin-$1/2$ R-parity odd axino $\ta$.
Typically in SUGRA one expects the saxion mass $m_s\sim m_{3/2}$ and the axino mass $m_{\ta}\alt m_{3/2}$.
The dark matter  is then comprised of two particles: the axion along with the LSP 
which is a Higgsino-like WIMP.
This is good news for natural SUSY since thermal higgsino-like WIMPs are typically underproduced
by a factor 10-15 below the measured dark matter abundance. The remainder can be comprised of axions.
 
The amount of dark matter generated in the early universe depends sensitively on the
properties of the axino and the saxion in addition to the SUSY spectrum and the axion.
For instance, thermally produced axinos can decay into LSPs after neutralino freeze-out 
thus augmenting the LSP abundance\cite{az1}. Saxions can be produced thermally or via coherent oscillations
(important at large $f_a$) and their decays can add to the LSP abundance, produce extra dark radiation
in the form of axions or dilute all relics via entropy production from decays to SM particles\cite{bbl}.
The calculation of the mixed axion-WIMP abundance requires solution of eight coupled Boltzmann
equations. Results from a mixed axion-higgsino dark matter calculation in natural SUSY are
shown in Fig. \ref{fig:sua1}\cite{dfsz2}. At low $f_a\sim 10^{10}$ GeV, then the thermal value of WIMP 
production is maintained since axinos decay before freeze-out. In this case the DM is axion-dominated\cite{bbc}.
For higher $f_a$ values, then axinos and saxions decay after freeze-out thus augmenting the 
WIMP abundance. For very large $f_a\agt 10^{14}$ GeV, then WIMPs are overproduced and those cases would be
excluded. Many of the high $f_a$ models are also excluded via violations of BBN constraints and by
overproduction of dark radiation- as parametrized by the effective number of extra neutrinos
in the universe $\Delta N_{eff}$.
\begin{figure}
\postscript{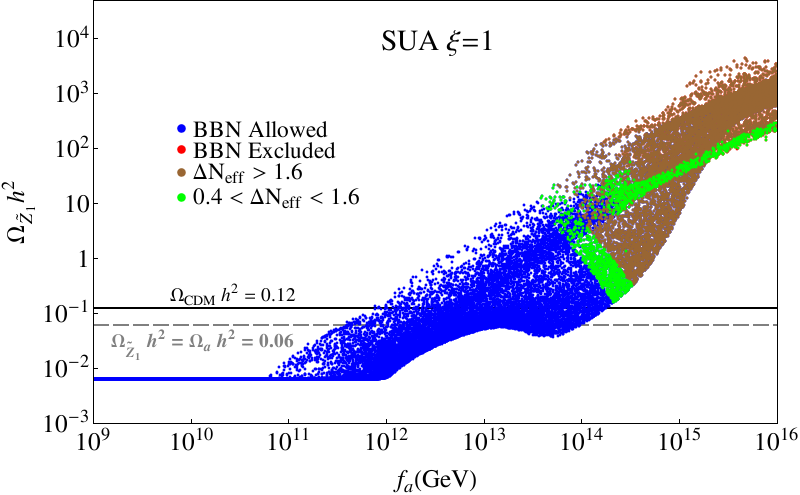}{0.95}
\caption{The neutralino relic density from a scan over
SUSY DFSZ parameter space for the RNS benchmark case 
labelled SUA with $\xi=1$.
The grey dashed line shows the points where DM consists of 50\% axions and
50\% neutralinos.
The red BBN-forbidden points occur at $f_a\agt 10^{14}$ GeV and are covered over by the brown 
$\Delta N_{eff}>1.6$ coloration. 
This latter region is excluded by Planck limits\cite{Planck:2015xua} of dark radiation
as parametrized by additional neutrino species beyond the SM value.
\label{fig:sua1}}
\end{figure}

As far as dark matter detection goes, WIMP production in RNS was examined in Ref. \cite{bbm}.
There, it is emphasized that the relevant theory prediction for WIMP direct detection is the
quantity $\xi\sigma^{SI} (\tz_1 p)$ where $\xi =\Omega_{\tz_1}h^2/0.12$ to reflect the
possibility that the WIMP local abundance may be highly depleted, and perhaps axion-dominated.
Nonetheless, WIMPs should be ultimately detected by ton-scale noble liquid detectors
because naturalness insures that the WIMP-Higgs coupling-- which is a product of 
higgsino and gaugino components-- is never small (see Fig. \ref{fig:SI}).
Prospects for indirect detection of higgsino-like WIMPs from WIMP-WIMP annilations 
to gamma rays or anti-matter are less lucrative since then the expected detection rates must
be scaled by $\xi^2$.
Meanwhile, we would also expect ultimate detection of axions if natural SUSY prevails\cite{axsearch}.
\begin{figure}[tbp]
\postscript{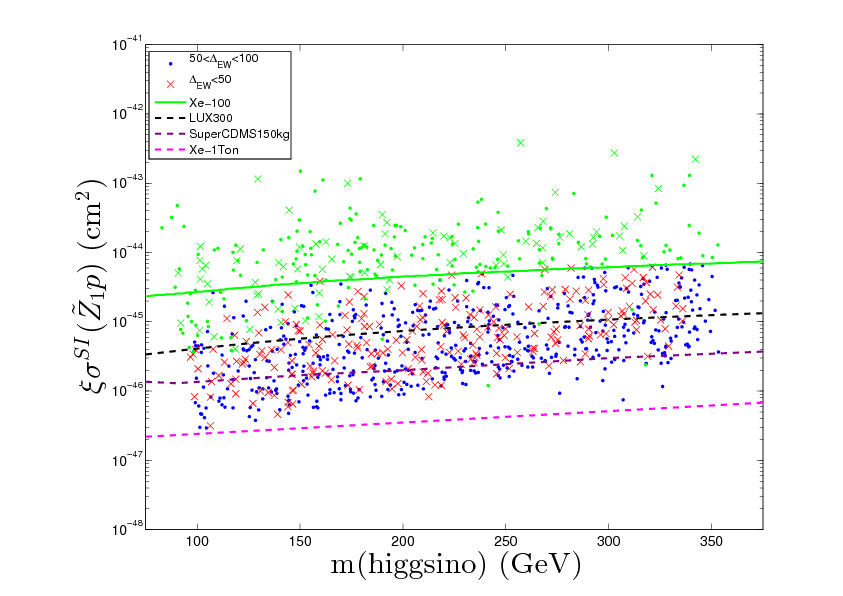}{1.0}
\caption{Plot of rescaled higgsino-like WIMP spin-independent 
direct detection rate $\xi \sigma^{SI}(\tz_1 p)$ 
versus $m(higgsino)$ from a scan over NUHM2 parameter space with $\Delta_{EW}<50$ (red crosses)
and $\Delta_{EW}<100$ (blue dots). 
Green points are excluded by current direct/indirect WIMP search experiments.
We also show the current reach from $Xe$-100 experiment, 
and projected reaches of LUX, SuperCDMS 150 kg and $Xe$-1 ton.
Plot from Ref. \cite{bbm}.
\label{fig:SI}}
\end{figure}

\section{Conclusions}

The framework of supergravity gauge theories with SUSY breaking taking place in a hidden sector
-- as put forth by Arnowitt-Chamseddine-Nath and others\cite{sugra} more than 30 years ago-- provides 
a compelling and elegant picture for physics beyond the Standard Model. 
SUGRA gauge theories allow for a solution to the naturalness/hierarchy problem, allow for the inclusion of 
gravity into particle physics and provide a candidate for cold dark matter. They receive indirect support
from the measured values of the gauge couplings, the top mass and the Higgs mass.
In spite of these successes, they have come under rather severe criticism of late due to
a (mis) perception of their increasing unnaturalness due to the rather high value of $m_h$ and due to 
increasingly severe search limits from LHC. 
Opinions have been voiced that we are witnessing the downfall
of one of the great paradigms of modern physics\cite{woit}.

In this paper, we have refuted this point of view. 
We noted that the oft-quoted, but seldom scrutinized, large-log measure
of naturalness neglects dependent terms which allow for large cancellations in the contributions to
the $Z$ or Higgs mass. It is time for this measure to be set aside: sub-TeV top squarks are not required
for SUSY naturalness.

The traditional BG measure of naturalness is almost always applied to the multi-parameter SUSY effective
theories where independent soft terms are introduced to parametrize a vast array of hidden sector possibilities.
If the soft terms of gravity-mediation are instead written as multiples of $m_{3/2}$, then their
{\it dependence} is explicitly displayed and their contributions to $m_Z$ or $m_h$ can be properly combined.
Thus, the BG measure is valid for SUGRA theories provided it is applied to Eq. \ref{eq:mzssugra}.
Once dependent terms are collected in their contributions to $m_Z$ or $m_h$, both large-log
and BG measures are seen to reduce to the electroweak measure $\Delta_{EW}$. 

The naturalness criterion for
low $\Delta_{EW}$ is that the higgsino mass $\mu$ and the weak scale soft term $|m_{H_u}|$ are not 
too far from $m_{Z,h}$-- in fact, the closer to $m_Z$ the better.
In addition, the top squarks can easily exist at the few TeV level so long as they are highly mixed by a large 
trilinear $A_t$ term. This condition also lifts the Higgs mass to $\sim 125$ GeV. These naturalness conditions
are easily realized in the two-parameter non-universal Higgs model where $m_{H_u}(GUT)$ is
typically about 30\% larger than $m_0$, the mass scale of the matter scalars. Then $m_{H_u}^2$ is radiatively 
driven to negative values rather close to $-m_Z^2$. Such SUSY models contain radiatively-driven naturalness
(RNS). Most other SUSY models which generate $m_h\sim 125$ GeV are found to be {\it un}-natural
(see Fig. \ref{fig:histo}\cite{dew}).
\begin{figure}[tbp]
\postscript{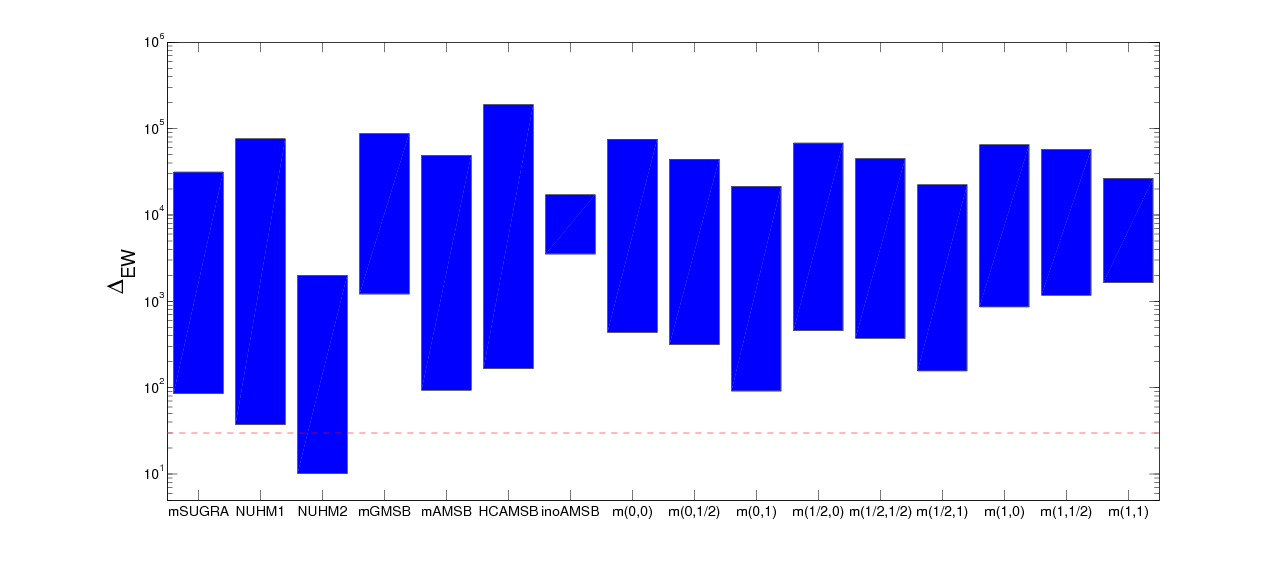}{1.15}
\caption{Histogram of range of $\Delta_{EW}$ values generated for each SUSY model
considered in Ref. \cite{dew}: mSUGRA, NUHM1,NUHM2,mGMSB,mAMSB,HCAMSB, inoAMSB and various 
versions of mirage-mediation for different modular weight choices. 
We would consider $\Delta_{EW}\alt 30$-- the lower the better--
as acceptable values for EW fine-tuning. This region is located below the dashed red line.
\label{fig:histo}}
\end{figure}

It is argued that naturalness should also be enforced in the QCD sector which leads to inclusion of
the invisible axion. In the DFSZ SUSY axion model, the SUSY $\mu$ problem is elegantly solved.
In fact, the $\mu$ term can itself be generated such that $\mu\ll m_{3/2}$ in a class
of DFSZ SUSY axion models with radiatively broken PQ symmetry.

For RNS SUSY models, SUSY might be accessible to LHC searches but could also easily evade LHC searches with
little cost to naturalness. The requisite light higgsino states, however, should be accessible to a linear
$e^+e^-$ collider operating with $\sqrt{s}>2m(higgsino)$.
For RNS SUSY, we expect ultimate detection of both a higgsino-like WIMP and the axion.
Discoveries such as these should vindicate the original vision put forth by Arnowitt-Chamseddine-Nath 
and others in their development of supergravity gauge theories.

Our ultimate plot is shown in Fig. \ref{fig:theory} where we show a figurative plot of 
theory space in the $1/natural$ vs. $1/simple$ plane.
The locus of the MSSM is along  a line at high simplicity but extending from 
highly natural to highly un-natural. 
SUSY models with radiatively-driven naturalness (RNS) lie in the highly simple and highly natural regime.
Future LHC searches will only be sensitive to a portion of the natural theory space.
An ILC $e^+e^-$ machine  will be required to test for the presence of the required light higgsino states.
\begin{figure}[tbp]
\postscript{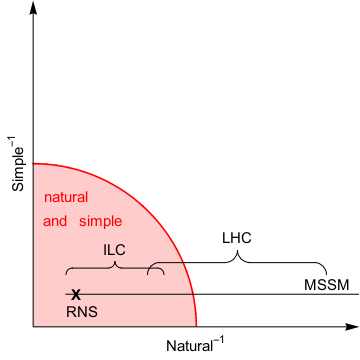}{1.05}
\caption{Figurative plot of theory space in the $1/natural$ vs. $1/simple$ plane
including the locus of the MSSM and the RNS SUSY models along with the
approximate reach of LHC and ILC.
\label{fig:theory}}
\end{figure}

{\it Acknowledgments:} 

This research was supported in part by grants from the United States Department of Energy.
We thank A. Mustafayev, P. Nath and X. Tata for comments on the manuscript.
We thank X. Tata, A. Mustafayev, P. Huang, W. Sreethawong, D. Mickelson,
M. Padeffke-Kirkland, K. J. Bae, A. Lessa and Hasan Serce for collaboration on the projects described herein.



\begin{thebibliography}{99}

\bibitem{atlas_h} G.~Aad {\it et al.}  [ATLAS Collaboration], 
 Phys. Lett. {\bf B716} (2012) 1.
%
\bibitem{cms_h} S.~Chatrchyan {\it et al.}  [CMS Collaboration],  
Phys. Lett. {\bf B716} (2012) 30.
%
\bibitem{techni1} L.~Susskind,
  Phys.\ Rev.\ D {\bf 20} (1979) 2619.
%
\bibitem{etc2} S.~Dimopoulos and L.~Susskind,
  Nucl.\ Phys.\ B {\bf 155} (1979) 237.
%
\bibitem{etc1} E.~Eichten and K.~D.~Lane,
  Phys.\ Lett.\ B {\bf 90} (1980) 125.
%
\bibitem{an} P.~Nath and R.~L.~Arnowitt,
  Phys.\ Lett.\ B {\bf 56} (1975) 177;
R.~L.~Arnowitt, P.~Nath and B.~Zumino,
  Phys.\ Lett.\ B {\bf 56} (1975) 81.
%
\bibitem{superg} D.~Z.~Freedman, P.~van Nieuwenhuizen and S.~Ferrara,
  Phys.\ Rev.\ D {\bf 13} (1976) 3214;
S.~Deser and B.~Zumino,
  Phys.\ Lett.\ B {\bf 62} (1976) 335;
S.~Ferrara, D.~Z.~Freedman, P.~van Nieuwenhuizen, P.~Breitenlohner, F.~Gliozzi and J.~Scherk,
  Phys.\ Rev.\ D {\bf 15} (1977) 1013.
%
\bibitem{Cremmer:1982en} E.~Cremmer, S.~Ferrara, L.~Girardello and A.~Van Proeyen,
  Nucl.\ Phys.\ B {\bf 212} (1983) 413;
see also 
P.~Nath, R.~L.~Arnowitt and A.~H.~Chamseddine,
  NUB-2613.
%
\bibitem{wessbagger} J.~Wess and J.~Bagger,
  Princeton, USA: Univ. Pr. (1992) 259 p.
%
\bibitem{wss} H.~Baer and X.~Tata,
  Cambridge, UK: Univ. Pr. (2006) 537 p.
%
\bibitem{sugra} H.~P.~Nilles,
  Phys.\ Lett.\ B {\bf 115} (1982) 193;
A.~Chamseddine, R.~Arnowitt and P.~Nath,
 Phys. Rev. Lett. {\bf 49} (1982) 970; 
R.~Barbieri, S.~Ferrara and   C.~Savoy, 
Phys. Lett. {\bf B119} (1982) 343; N.~Ohta,
  Prog.~Theor.~Phys.~{\bf 70}, 542 (1983); L.~Hall, J.~Lykken and
  S.~Weinberg, Phys. Rev. {\bf D27} (1983) 2359;
for an early review, see {\it e.g.} H.~P.~Nilles,
  Phys.\ Rept.\  {\bf 110} (1984) 1; for a recent review, see
  D.~J.~H.~Chung, L.~L.~Everett, G.~L.~Kane, S.~F.~King, J.~D.~Lykken and L.~T.~Wang,
  Phys.\ Rept.\  {\bf 407} (2005) 1.
%
\bibitem{acnreview}R.~Arnowitt, A.~H.~Chamseddine and P.~Nath,
  Int.\ J.\ Mod.\ Phys.\ A {\bf 27} (2012) 1230028; see also
P.~Nath,
  arXiv:1502.00639 [physics.hist-ph];
H.~P.~Nilles,
  Nucl.\ Phys.\ Proc.\ Suppl.\  {\bf 101} (2001) 237.
%
\bibitem{Soni:1983rm} S.~K.~Soni and H.~A.~Weldon,
  Phys.\ Lett.\ B {\bf 126} (1983) 215.
%
\bibitem{kl}V.~S.~Kaplunovsky and J.~Louis,
  Phys.\ Lett.\ B {\bf 306} (1993) 269. 
%
\bibitem{Brignole:1993dj} A.~Brignole, L.~E.~Ibanez and C.~Munoz,
  Nucl.\ Phys.\ B {\bf 422} (1994) 125
   [Erratum-ibid.\ B {\bf 436} (1995) 747];
A.~Brignole, L.~E.~Ibanez and C.~Munoz,
  Adv.\ Ser.\ Direct.\ High Energy Phys.\  {\bf 21} (2010) 244
  [hep-ph/9707209].
%
\bibitem{gaugino} S.~Ferrara, L.~Girardello and H.~P.~Nilles,
  Phys.\ Lett.\ B {\bf 125} (1983) 457.
%
\bibitem{RGEs} S.~Dimopoulos, S.~Raby and F.~Wilczek,
  Phys.\ Rev.\ D {\bf 24} (1981) 1681;
K.~Inoue, A.~Kakuto, H.~Komatsu and S.~Takeshita,
  Prog.\ Theor.\ Phys.\  {\bf 68} (1982) 927
   [Erratum-ibid.\  {\bf 70} (1983) 330]
   [Prog.\ Theor.\ Phys.\  {\bf 70} (1983) 330];
L.~Alvarez-Gaume, J.~Polchinski and M.~B.~Wise,
  Nucl.\ Phys.\ B {\bf 221} (1983) 495.
K.~Inoue, A.~Kakuto, H.~Komatsu and S.~Takeshita,
  Prog.\ Theor.\ Phys.\  {\bf 71} (1984) 413.
%
\bibitem{rge1992}
G.~G.~Ross and R.~G.~Roberts,
  Nucl.\ Phys.\ B {\bf 377} (1992) 571;
R.~L.~Arnowitt and P.~Nath,
  Phys.\ Rev.\ Lett.\  {\bf 69} (1992) 725;
  V.~D.~Barger, M.~S.~Berger and P.~Ohmann,
  Phys.\ Rev.\ D {\bf 47} (1993) 1093;
V.~D.~Barger, M.~S.~Berger and P.~Ohmann,
  Phys.\ Rev.\ D {\bf 49} (1994) 4908;
S.~P.~Martin and M.~T.~Vaughn,
  Phys.\ Rev.\ D {\bf 50} (1994) 2282
   [Erratum-ibid.\ D {\bf 78} (2008) 039903].
%
\bibitem{Dimopoulos:1981zb}
  S.~Dimopoulos and H.~Georgi,
  Nucl.\ Phys.\ B {\bf 193} (1981) 150.
%
\bibitem{polonsky} N.~Polonsky and A.~Pomarol,
  Phys.\ Rev.\ D {\bf 51} (1995) 6532.
%
\bibitem{bdqt} H.~Baer, M.~A.~Diaz, P.~Quintana and X.~Tata,
  JHEP {\bf 0004} (2000) 016.
%
\bibitem{pbmz} D.~M.~Pierce, J.~A.~Bagger, K.~T.~Matchev and R.~j.~Zhang,
  Nucl.\ Phys.\ B {\bf 491} (1997) 3. 
%
\bibitem{isajet} ISAJET, by H.~Baer, F.~Paige, S.~Protopopescu and
X.~Tata, hep-ph/0312045.
%
\bibitem{Allanach:2003jw} B.~C.~Allanach, S.~Kraml and W.~Porod,
  JHEP {\bf 0303} (2003) 016.
%
\bibitem{rewsb} L. E. Iba\~nez and G. G. Ross, Phys. Lett. {\bf B110}, 215
(1982); K. Inoue {\it et al.} Prog. Theor. Phys. {\bf 68}, 927 (1982)
and {\bf 71}, 413 (1984); 
L.~Iba\~nez, Phys. Lett. {\bf B118}, 73 (1982); 
 H.~P.~Nilles, M.~Srednicki and D.~Wyler,
  Phys.\ Lett.\ B {\bf 120} (1983) 346;
J.~Ellis, J.~Hagelin, D.~Nanopoulos and M.~Tamvakis,
Phys. Lett. {\bf B125}, 275 (1983); 
L.~Alvarez-Gaum\'e. J.~Polchinski and M.~Wise,
Nucl. Phys. {\bf B221}, 495 (1983);
B.~A.~Ovrut and S.~Raby,
  Phys.\ Lett.\ B {\bf 130} (1983) 277;
for a review, see 
L.~E.~Ibanez and G.~G.~Ross,
  Comptes Rendus Physique {\bf 8} (2007) 1013.
%
\bibitem{GM} G.~F.~Giudice and A.~Masiero,
  Phys.\ Lett.\ B {\bf 206} (1988) 480.
%
\bibitem{nmssm} For a review, see U.~Ellwanger, C.~Hugonie and A.~M.~Teixeira,
  Phys.\ Rept.\  {\bf 496} (2010) 1.
%
\bibitem{KN} J.~E.~Kim and H.~P.~Nilles,
  Phys.\ Lett.\ B {\bf 138} (1984) 150.
%
\bibitem{Lee:1977eg}
D.~A.~Dicus and V.~S.~Mathur,
  Phys.\ Rev.\ D {\bf 7} (1973) 3111;
  B.~W.~Lee, C.~Quigg and H.~B.~Thacker,
  Phys.\ Rev.\ D {\bf 16} (1977) 1519.
%
\bibitem{mhiggs} M.~S.~Carena and H.~E.~Haber,
Prog.\ Part.\ Nucl.\ Phys.\  {\bf 50} (2003) 63.
%
\bibitem{h125}  H.~Baer, V.~Barger and A.~Mustafayev,
  Phys.\ Rev.\ D {\bf 85} (2012) 075010.
%
\bibitem{cascade}H.~Baer, J.~R.~Ellis, G.~B.~Gelmini, D.~V.~Nanopoulos and X.~Tata,
  Phys.\ Lett.\ B {\bf 161} (1985) 175;
G.~Gamberini,
  Z.\ Phys.\ C {\bf 30} (1986) 605;
H.~Baer, V.~D.~Barger, D.~Karatas and X.~Tata,
  Phys.\ Rev.\ D {\bf 36} (1987) 96;
H.~Baer, R.~M.~Barnett, M.~Drees, J.~F.~Gunion, H.~E.~Haber, D.~L.~Karatas and X.~R.~Tata,
  Int.\ J.\ Mod.\ Phys.\ A {\bf 2} (1987) 1131;
R.~M.~Barnett, J.~F.~Gunion and H.~E.~Haber,
  Phys.\ Rev.\ D {\bf 37} (1988) 1892.
H.~Baer, A.~Bartl, D.~Karatas, W.~Majerotto and X.~Tata,
  Int.\ J.\ Mod.\ Phys.\ A {\bf 4} (1989) 4111;
H.~Baer, X.~Tata and J.~Woodside,
  Phys.\ Rev.\ D {\bf 42} (1990) 1568;
A.~Bartl, W.~Majerotto, B.~Mosslacher, N.~Oshimo and S.~Stippel,
  Phys.\ Rev.\ D {\bf 43} (1991) 2214;
A.~Bartl, W.~Majerotto and W.~Porod,
  Z.\ Phys.\ C {\bf 64} (1994) 499.
 
%
\bibitem{bcpt} H.~Baer, X.~Tata and J.~Woodside,
  Phys.\ Rev.\ D {\bf 45} (1992) 142; 
H.~Baer, C.~h.~Chen, F.~Paige and X.~Tata,
  Phys.\ Rev.\ D {\bf 53} (1996) 6241.
%
\bibitem{atlas_susy}
G.~Aad {\it et al.}  [ATLAS Collaboration],
  JHEP {\bf 1409} (2014) 176; 
G.~Aad {\it et al.}  [ATLAS Collaboration],
  arXiv:1501.03555 [hep-ex].
%
\bibitem{cms_susy}
CMS Collaboration [CMS Collaboration],
  CMS-PAS-SUS-12-016.
%
\bibitem{pdb} K.~A.~Olive {\it et al.}  [Particle Data Group Collaboration],
  Chin.\ Phys.\ C {\bf 38} (2014) 090001.
%
\bibitem{Shifman:2012na}
  M.~Shifman,
  Mod.\ Phys.\ Lett.\ A {\bf 27} (2012) 1230043.
%
\bibitem{Barbieri:2013vca}
  R.~Barbieri,
  Phys.\ Scripta T {\bf 158} (2013) 014006.
%
\bibitem{Giudice:2013yca}
  G.~F.~Giudice,
  PoS EPS {\bf -HEP2013} (2013) 163.
%
\bibitem{Altarelli:2014roa}
  G.~Altarelli,
  EPJ Web Conf.\  {\bf 71} (2014) 00005.
%
\bibitem{Craig:2013cxa}
  N.~Craig,
  arXiv:1309.0528 [hep-ph].
%
\bibitem{Murayama:2014ita}
  H.~Murayama,
  Phys.\ Scripta T {\bf 158} (2013) 014025
%
\bibitem{Ross:2014mua}
  G.~G.~Ross,
  Eur.\ Phys.\ J.\ C {\bf 74} (2014) 2699.
%
\bibitem{Lykken:2014bca}
  J.~Lykken and M.~Spiropulu,
  Sci.\ Am.\  {\bf 310N5} (2014) 5,  36.
%
\bibitem{Dine:2015xga}
  M.~Dine,
  arXiv:1501.01035 [hep-ph].
%
\bibitem{barbstrum} R.~Barbieri and A.~Strumia,
  Phys.\ Lett.\ B {\bf 433} (1998) 63;
A.~Birkedal, Z.~Chacko and M.~K.~Gaillard,
  JHEP {\bf 0410} (2004) 036; R.~Dermisek and J.~F.~Gunion,
  Phys.\ Rev.\ Lett.\  {\bf 95} (2005) 041801;
K.~Choi, K.~S.~Jeong, T.~Kobayashi and K.~i.~Okumura,
  Phys.\ Lett.\ B {\bf 633} (2006) 355;
S.-G.~Kim, N.~Maekawa, A.~Matsuzaki, K.~Sakurai, A.~I.~Sanda and T.~Yoshikawa,
  Phys.\ Rev.\ D {\bf 74} (2006) 115016;
K.~Choi, K.~S.~Jeong, T.~Kobayashi and K.~i.~Okumura,
  Phys.\ Rev.\ D {\bf 75} (2007) 095012;
B.~Dutta and Y.~Mimura,
  Phys.\ Lett.\ B {\bf 648} (2007) 357;
B.~Dutta, Y.~Mimura and D.~V.~Nanopoulos,
  Phys.\ Lett.\ B {\bf 656} (2007) 199;
K.~S.~Babu, I.~Gogoladze, M.~U.~Rehman and Q.~Shafi,
  Phys.\ Rev.\ D {\bf 78} (2008) 055017;
B.~Bellazzini, C.~Csaki, A.~Delgado and A.~Weiler,
  Phys.\ Rev.\ D {\bf 79} (2009) 095003;
A.~Delgado, C.~Kolda, J.~P.~Olson and A.~de la Puente,
  Phys.\ Rev.\ Lett.\  {\bf 105} (2010) 091802;
T.~Gherghetta, B.~von Harling and N.~Setzer,
  JHEP {\bf 1107} (2011) 011;
D.~Feldman, G.~Kane, E.~Kuflik and R.~Lu,
  Phys.\ Lett.\ B {\bf 704} (2011) 56;
J.~E.~Younkin and S.~P.~Martin,
  Phys.\ Rev.\ D {\bf 85} (2012) 055028.
%
\bibitem{radpq} K.~J.~Bae, H.~Baer and H.~Serce,
  Phys.\ Rev.\ D {\bf 91} (2015) 015003.
%
\bibitem{comp} H.~Baer, V.~Barger and D.~Mickelson,
  Phys.\ Rev.\ D {\bf 88} (2013) 095013.
%
\bibitem{azar_xt} A.~Mustafayev and X.~Tata,
  Indian J.\ Phys.\  {\bf 88} (2014) 991.
%
\bibitem{dew} H.~Baer, V.~Barger, D.~Mickelson and M.~Padeffke-Kirkland,
  Phys.\ Rev.\ D {\bf 89} (2014) 115019.
%
\bibitem{kane} G.~Kane, C.~Kolda, L.~Roszkowski and J.~Wells, \prd{49}{1994}{6173}.
%
\bibitem{ac1} G.~W.~Anderson and D.~J.~Castano,
  \plb{347}{1995}{300} and   \prd{52}{1995}{1693}.
%
\bibitem{dg} S.~Dimopoulos and G.~F.~Giudice,
  \plb{357}{1995}{573}.
%
\bibitem{ccn} 
S.~Akula, M.~Liu, P.~Nath and G.~Peim,
  Phys.\ Lett.\ B {\bf 709} (2012) 192;
M.~Liu and P.~Nath,
  Phys.\ Rev.\ D {\bf 87} (2013) 095012.
%
\bibitem{ellis2} P.~H.~Chankowski, J.~R.~Ellis and S.~Pokorski,
  \plb{423}{1998}{327};
P.~H.~Chankowski, J.~R.~Ellis, M.~Olechowski and S.~Pokorski,
  \npb{544}{1999}{39}.
%
\bibitem{king}
G.~L.~Kane and S.~F.~King,
  \plb{451}{1999}{113};
M.~Bastero-Gil, G.~L.~Kane and S.~F.~King,
  \plb{474}{2000}{103}.
%
\bibitem{casas} J.~A.~Casas, J.~R.~Espinosa and I.~Hidalgo,
  JHEP {\bf 0401} (2004) 008.
%
\bibitem{fp}
  J.~L.~Feng, K.~T.~Matchev and T.~Moroi,
  Phys.\ Rev.\ D {\bf 61} (2000) 075005;
J.~L.~Feng, K.~T.~Matchev and T.~Moroi,
  hep-ph/0003138;
J.~L.~Feng and D.~Sanford,
  Phys.\ Rev.\ D {\bf 86} (2012) 055015.
%
\bibitem{Harnik:2003rs}
  R.~Harnik, G.~D.~Kribs, D.~T.~Larson and H.~Murayama,
  Phys.\ Rev.\ D {\bf 70} (2004) 015002.
%
\bibitem{Nomura:2005qg}
Y.~Nomura and B.~Tweedie,
  Phys.\ Rev.\ D {\bf 72} (2005) 015006;
Y.~Nomura, D.~Poland and B.~Tweedie,
  Nucl.\ Phys.\ B {\bf 745} (2006) 29.
%
\bibitem{Athron:2007ry}
  P.~Athron and D.~J.~Miller,
  Phys.\ Rev.\ D {\bf 76} (2007) 075010.
%
\bibitem{ross} S.~Cassel, D.~M.~Ghilencea and G.~G.~Ross,
  \npb{825}{2010}{203} and
  \npb{835}{2010}{110};
S.~Cassel, D.~M.~Ghilencea, S.~Kraml, A.~Lessa and G.~G.~Ross,
  \jhep{1105}{2011}{120};
G.~G.~Ross and K.~Schmidt-Hoberg,
  Nucl.\ Phys.\ B {\bf 862} (2012) 710;
G.~G.~Ross, K.~Schmidt-Hoberg and F.~Staub,
  JHEP {\bf 1208} (2012) 074;
D.~M.~Ghilencea and G.~G.~Ross,
  Nucl.\ Phys.\ B {\bf 868} (2013) 65;
A.~Kaminska, G.~G.~Ross and K.~Schmidt-Hoberg,
  JHEP {\bf 1311} (2013) 209.
%
\bibitem{derm_kim} R.~Dermisek and H.~D.~Kim,
  Phys.\ Rev.\ Lett.\  {\bf 96} (2006) 211803.
%
\bibitem{Allanach:2012vj}
  B.~C.~Allanach and B.~Gripaios,
  JHEP {\bf 1205} (2012) 062.
%
\bibitem{shafi} 
  I.~Gogoladze, F.~Nasir and Q.~Shafi,
  Int.\ J.\ Mod.\ Phys.\ A {\bf 28}, 1350046 (2013);
I.~Gogoladze, F.~Nasir and Q.~Shafi,
  JHEP {\bf 1311} (2013) 173.
%
\bibitem{perel} M.~Perelstein and B.~Shakya,
  JHEP {\bf 1110} (2011) 142; 
M.~Perelstein and B.~Shakya,
  Phys.\ Rev.\ D {\bf 88} (2013) 075003.
%
\bibitem{Wymant:2012zp}
  C.~Wymant,
  Phys.\ Rev.\ D {\bf 86} (2012) 115023.
%
\bibitem{antusch} S.~Antusch, L.~Calibbi, V.~Maurer, M.~Monaco and M.~Spinrath,
  Phys.\ Rev.\ D {\bf 85} (2012) 035025 and JHEP {\bf 1301} (2013) 187. 
%
\bibitem{Kribs:2013lua}
  G.~D.~Kribs, A.~Martin and A.~Menon,
  Phys.\ Rev.\ D {\bf 88} (2013) 035025.
%
\bibitem{hardy} E.~Hardy,
  JHEP {\bf 1310} (2013) 133;
E.~Hardy,
  JHEP {\bf 1403} (2014) 069.
%
\bibitem{Galloway:2013dma}
  J.~Galloway, M.~A.~Luty, Y.~Tsai and Y.~Zhao,
  Phys.\ Rev.\ D {\bf 89} (2014) 075003.
%
\bibitem{Craig:2013fga}
  N.~Craig and K.~Howe,
  JHEP {\bf 1403} (2014) 140.
%
\bibitem{sug19}
H.~Baer, V.~Barger and M.~Padeffke-Kirkland,
  Phys.\ Rev.\ D {\bf 88} (2013) 055026.
%
\bibitem{Fichet:2012sn}
  S.~Fichet,
  Phys.\ Rev.\ D {\bf 86} (2012) 125029.
%
\bibitem{Younkin:2012ui}
  J.~E.~Younkin and S.~P.~Martin,
  Phys.\ Rev.\ D {\bf 85} (2012) 055028.
%
\bibitem{Kowalska:2013ica} K.~Kowalska and E.~M.~Sessolo,
  Phys.\ Rev.\ D {\bf 88} (2013) 075001.
%
\bibitem{han} C.~Han, K.~-i.~Hikasa, L.~Wu, J.~M.~Yang and Y.~Zhang,
  JHEP {\bf 1310} (2013) 216.
%
\bibitem{Krippendorf:2012ir}
  S.~Krippendorf, H.~P.~Nilles, M.~Ratz and M.~W.~Winkler,
  Phys.\ Lett.\ B {\bf 712} (2012) 87.
%
\bibitem{Dudas:2013pja}
 E.~Dudas, G.~von Gersdorff, S.~Pokorski and R.~Ziegler,
  JHEP {\bf 1401} (2014) 117.
%
\bibitem{Arvanitaki:2013yja}
A.~Arvanitaki, M.~Baryakhtar, X.~Huang, K.~van Tilburg and G.~Villadoro,
  JHEP {\bf 1403} (2014) 022.
%
\bibitem{Fan:2014txa}
  J.~Fan and M.~Reece,
  arXiv:1401.7671 [hep-ph].
%
\bibitem{Gherghetta:2014xea}
T.~Gherghetta, B.~von Harling, A.~D.~Medina and M.~A.~Schmidt,
  JHEP {\bf 1404} (2014) 180.
%
\bibitem{Kowalska:2014hza}
K.~Kowalska, L.~Roszkowski, E.~M.~Sessolo and S.~Trojanowski,
  JHEP {\bf 1404} (2014) 166.
%
\bibitem{Martin:2013aha}
  S.~P.~Martin,
  Phys.\ Rev.\ D {\bf 89} (2014) 035011.
%
\bibitem{Fowlie:2014xha}
A.~Fowlie,
  Phys.\ Rev.\ D {\bf 90} (2014) 015010.
%
\bibitem{Casas:2014eca}
  J.~A.~Casas, J.~M.~Moreno, S.~Robles, K.~Rolbiecki and B.~Zaldivar,
  arXiv:1407.6966 [hep-ph].
%
\bibitem{ltr}
H.~Baer, V.~Barger, P.~Huang, A.~Mustafayev and X.~Tata,
Phys. Rev. Lett. {\bf 109} (2012) 161802.
%
\bibitem{rns} H.~Baer, V.~Barger, P.~Huang, D.~Mickelson, A.~Mustafayev and X.~Tata,
  Phys.\ Rev.\ D {\bf 87} (2013) 115028.
%
\bibitem{feng} For a recent review, see {\it e.g.}
J.~L.~Feng,
  Ann.\ Rev.\ Nucl.\ Part.\ Sci.\  {\bf 63} (2013) 351.
%
\bibitem{deg} H.~Baer, V.~Barger, M.~Padeffke-Kirkland and X.~Tata,
  Phys.\ Rev.\ D {\bf 89} (2014) 037701.
%
\bibitem{Chan:1997bi}
  K.~L.~Chan, U.~Chattopadhyay and P.~Nath,
  Phys.\ Rev.\ D {\bf 58} (1998) 096004.
%
\bibitem{hgsno} H.~Baer, V.~Barger and P.~Huang,
  JHEP {\bf 1111} (2011) 031.
%
\bibitem{Barbieri:2009ev}
  R.~Barbieri and D.~Pappadopulo,
  JHEP {\bf 0910} (2009) 061.
%
\bibitem{Cohen:2015ala}
  T.~Cohen, J.~Kearney and M.~Luty,
  arXiv:1501.01962 [hep-ph].
%
\bibitem{Nelson:2015cea}
  A.~E.~Nelson and T.~S.~Roy,
  arXiv:1501.03251 [hep-ph].
%
\bibitem{nuhm2}  D.~Matalliotakis and H.~P.~Nilles,
  Nucl.\ Phys.\ B {\bf 435} (1995) 115;
P.~Nath and R.~L.~Arnowitt,
  Phys.\ Rev.\ D {\bf 56} (1997) 2820;
J. Ellis, K. Olive and Y. Santoso, Phys. Lett. {\bf B539} (2002) 107;
J. Ellis, T. Falk, K. Olive and Y. Santoso, 
Nucl. Phys. {\bf B652} (2003) 259;
H.~Baer, A.~Mustafayev, S.~Profumo, A.~Belyaev and X. Tata, 
JHEP{\bf 0507} (2005) 065.
%
\bibitem{kn} R.~Kitano and Y.~Nomura,
  Phys.\ Lett.\ B {\bf 631} (2005) 58;
R.~Kitano and Y.~Nomura,
  Phys.\ Rev.\ D {\bf 73} (2006) 095004.
%
\bibitem{papucci} M.~Papucci, J.~T.~Ruderman and A.~Weiler,
  JHEP {\bf 1209} (2012) 035.
%
\bibitem{brust} C.~Brust, A.~Katz, S.~Lawrence and R.~Sundrum,
  JHEP {\bf 1203} (2012) 103.
%
\bibitem{Evans:2013jna} J.~A.~Evans, Y.~Kats, D.~Shih and M.~J.~Strassler,
  JHEP {\bf 1407} (2014) 101.
%
\bibitem{Ellis:1986yg}
  J.~R.~Ellis, K.~Enqvist, D.~V.~Nanopoulos and F.~Zwirner,
  Mod.\ Phys.\ Lett.\ A {\bf 1} (1986) 57.
%
\bibitem{bg}
  R.~Barbieri and G.~F.~Giudice,
  Nucl.\ Phys.\ B {\bf 306} (1988) 63.
%
\bibitem{munoz} L.~E.~Ibanez, C.~Lopez and C.~Munoz,
  Nucl. Phys. {\bf B256} (1985) 218;
A.~Lleyda and C.~Munoz,
  Phys. Lett. {\bf B317} (1993) 82.
%
\bibitem{abe} H.~Abe, T.~Kobayashi and Y.~Omura,
  Phys. Rev. {\bf D76} (2007) 015002.
%
\bibitem{martin}  S.~P.~Martin,
  Phys. Rev. {\bf D75} (2007) 115005.
%
\bibitem{rnshiggs} K.~J.~Bae, H.~Baer, V.~Barger, D.~Mickelson and M.~Savoy,
  Phys.\ Rev.\ D {\bf 90} (2014) 075010.
%
\bibitem{U1} S.~Weinberg,
  Phys.\ Rev.\ D {\bf 11} (1975) 3583. 
%
\bibitem{tHooft}G.~'t Hooft,
  Phys.\ Rev.\ Lett.\  {\bf 37} (1976) 8. 
%
\bibitem{pqww}
  R.~D.~Peccei and H.~R.~Quinn,
  Phys.\ Rev.\ Lett.\  {\bf 38}, 1440 (1977);
 S. Weinberg, Phys.\ Rev.\ Lett.\  {\bf 40} (1978) 223;
 F. Wilczek, Phys. Rev. lett. {\bf 40} (1978) 279.
%
\bibitem{ksvz}
 J. E. Kim, Phys. Rev. Lett. {\bf 43} (1979) 103;
 M. A. Shifman, A. Vainstein and V. I. Zakharov,
 Nucl. Phys. {\bf B166}( 1980) 493.
%
\bibitem{dfsz}
M. Dine, W. Fischler and M. Srednicki,
Phys. Lett. {\bf B104} (1981) 199;
A. P. Zhitnitskii, Sov. J. Phys. {\bf 31} (1980) 260.
%
\bibitem{axdm}  L. F. Abbott and P. Sikivie, Phys. Lett. {\bf B120} (1983) 133;
J. Preskill, M. Wise and F. Wilczek, Phys. Lett. {\bf B120} (1983) 127;
M. Dine and W. Fischler, Phys. Lett. {\bf B120} (1983) 137;
M. Turner, Phys. Rev. {\bf D33} (1986) 889.
%
\bibitem{susydfsz} E.~J.~Chun,
  Phys.\ Rev.\ D {\bf 84} (2011) 043509; K.~J.~Bae, E.~J.~Chun and S.~H.~Im,
  JCAP {\bf 1203} (2012) 013; 
%
\bibitem{msy} H.~Murayama, H.~Suzuki and T.~Yanagida,
  Phys.\ Lett.\ B {\bf 291} (1992) 418;
T.~Gherghetta and G.~L.~Kane,
  Phys.\ Lett.\ B {\bf 354} (1995) 300;
K.~Choi, E.~J.~Chun and J.~E.~Kim,
  Phys.\ Lett.\ B {\bf 403} (1997) 209.
%
\bibitem{nugm} H.~Baer, V.~Barger, P.~Huang, D.~Mickelson, M.~Padeffke-Kirkland and X.~Tata,
  arXiv:1501.06357 [hep-ph].
%
\bibitem{dine} M.~Dine, A.~Kagan and S.~Samuel,
  Phys.\ Lett.\ B {\bf 243} (1990) 250;
A.~G.~Cohen, D.~B.~Kaplan and A.~E.~Nelson,
  Phys.\ Lett.\ B {\bf 388} (1996) 588;
J.~Bagger, J.~L.~Feng and N.~Polonsky,
  Nucl.\ Phys.\ B {\bf 563} (1999) 3;
T.~Moroi and M.~Nagai,
  Phys.\ Lett.\ B {\bf 723} (2013) 107.
%
\bibitem{gltotop} H.~Baer, X.~Tata and J.~Woodside,
  Phys.\ Rev.\ D {\bf 42} (1990) 1568;
%
\bibitem{btw2} V.~D.~Barger, W.~Y.~Keung and R.~J.~N.~Phillips,
  Phys.\ Rev.\ Lett.\  {\bf 55} (1985) 166;
H.~Baer, X.~Tata and J.~Woodside,
  Phys.\ Rev.\ D {\bf 45} (1992) 142; R.~M.~Barnett, J.~F.~Gunion and H.~E.~Haber,
  Phys.\ Lett.\ B {\bf 315} (1993) 349.
%
\bibitem{lhc} H.~Baer, V.~Barger, P.~Huang, D.~Mickelson, A.~Mustafayev, W.~Sreethawong and X.~Tata,
  JHEP {\bf 1312} (2013) 013.
%
\bibitem{chan} C.~Han, A.~Kobakhidze, N.~Liu, A.~Saavedra, L.~Wu and J.~M.~Yang,
  JHEP {\bf 1402} (2014) 049.
%
\bibitem{monojet} H.~Baer, A.~Mustafayev and X.~Tata,
  Phys.\ Rev.\ D {\bf 89} (2014) 055007.
%
\bibitem{kribs} Z.~Han, G.~D.~Kribs, A.~Martin and A.~Menon,
  Phys.\ Rev.\ D {\bf 89} (2014) 075007.
%
\bibitem{dilep} H.~Baer, A.~Mustafayev and X.~Tata,
  Phys.\ Rev.\ D {\bf 90} (2014) 115007.
%
\bibitem{lhcltr} H.~Baer, V.~Barger, P.~Huang, D.~Mickelson, A.~Mustafayev, W.~Sreethawong and X.~Tata,
  Phys.\ Rev.\ Lett.\  {\bf 110} (2013) 151801.
%
\bibitem{ilc} H.~Baer, V.~Barger, D.~Mickelson, A.~Mustafayev and X.~Tata,
  JHEP {\bf 1406} (2014) 172.
%
\bibitem{az1} K.~Y.~Choi, J.~E.~Kim, H.~M.~Lee and O.~Seto,
  Phys.\ Rev.\ D {\bf 77} (2008) 123501;
H.~Baer, A.~Lessa, S.~Rajagopalan and W.~Sreethawong,
  JCAP {\bf 1106} (2011) 031.
%
\bibitem{bbl}K.~J.~Bae, H.~Baer and A.~Lessa,
  JCAP {\bf 1304} (2013) 041. 
%
\bibitem{dfsz2} K.~J.~Bae, H.~Baer, A.~Lessa and H.~Serce,
  JCAP {\bf 1410} (2014) 10,  082.
%
\bibitem{bbc} K.~J.~Bae, H.~Baer and E.~J.~Chun,
  Phys.\ Rev.\ D {\bf 89} (2014) 031701;
K.~J.~Bae, H.~Baer and E.~J.~Chun,
  JCAP {\bf 1312} (2013) 028.
%
\bibitem{Planck:2015xua}
  P.~A.~R.~Ade {\it et al.}  [Planck Collaboration],
  arXiv:1502.01589 [astro-ph.CO].
%
\bibitem{bbm} H.~Baer, V.~Barger and D.~Mickelson,
  Phys.\ Lett.\ B {\bf 726} (2013) 330; 
H.~Baer, V.~Barger, D.~Mickelson and X.~Tata,
  arXiv:1306.4183 [hep-ph].
%
\bibitem{axsearch}
L.\ Duffy, {\it et.~al.}, \prl{95}{2005}{091304} and \prd{74}{2006}{012006};
for a review, see S.~J.~Asztalos, L. Rosenberg, K. van Bibber, P. Sikivie
and K. Zioutas,
  {\it Ann.\ Rev.\ Nucl.\ Part.\ Sci.\ } {\bf 56} (2006) 293.
%
\bibitem{woit} Peter Woit, April 15, 2014 post 
in \verb|http://www.math.columbia.edu/~woit/wordpress/| .
%


\end{thebibliography}
\end{document}